\documentclass[reprint,
superscriptaddress,
nofootinbib,
amsmath,amssymb,
aps,
prd]{revtex4-2}

\usepackage{graphicx}

\usepackage[lofdepth,lotdepth,caption=false]{subfig}
\usepackage{dcolumn}
\usepackage{bm}
\usepackage{multirow}
\usepackage{siunitx}
\usepackage{xcolor}
\usepackage{comment}

\definecolor{cerulean}{rgb}{0., 0.62,0.9}

\newcommand{\meV}{{\rm meV}}
\newcommand{\eV}{{\rm eV}}
\newcommand{\keV}{{\rm keV}}
\newcommand{\MeV}{{\rm MeV}}

\newcommand{\bfq}{{\bf q}}
\newcommand{\bfk}{{\bf k}}
\newcommand{\bfG}{{\bf G}}
\newcommand{\us}{{$\rm\mu s$}}

\usepackage{hyperref}
\hypersetup{
     colorlinks   = true,
     citecolor    = blue,
	 linkcolor=blue
}

\begin{document}

\title{SiC Detectors for Sub-GeV Dark Matter}

\author{Sin\'{e}ad M. Griffin}\affiliation{
Materials Sciences Division, Lawrence Berkeley National Laboratory, Berkeley, CA 94720, USA}\affiliation{Molecular Foundry, Lawrence Berkeley National Laboratory, Berkeley, CA 94720, USA}
\author{Yonit Hochberg}\affiliation{Racah Institute of Physics, Hebrew University of Jerusalem, Jerusalem 91904, Israel}
\author{Katherine Inzani}\affiliation{
Materials Sciences Division, Lawrence Berkeley National Laboratory, Berkeley, CA 94720, USA}\affiliation{Molecular Foundry, Lawrence Berkeley National Laboratory, Berkeley, CA 94720, USA}
\author{Noah Kurinsky}\affiliation{Fermi National Accelerator Laboratory, Batavia, IL 60510, USA} \affiliation{Kavli Institute for Cosmological Physics, University of Chicago, Chicago, IL 60637, USA}
\author{Tongyan Lin}\affiliation{Department of Physics, University of California, San Diego, California 92093, USA}
\author{To Chin Yu}\affiliation{Department of Physics, Stanford University, Stanford, CA 94305, USA}
\affiliation{SLAC National Accelerator Laboratory, 2575 Sand Hill Road, Menlo Park, CA 94025, USA}

\date{\today}

\begin{abstract}
We propose the use of silicon carbide (SiC) for direct detection of sub-GeV dark matter. SiC has properties similar to both silicon and diamond, but has two key advantages: (i)~it is a polar semiconductor which allows sensitivity to a broader range of  dark matter candidates; and (ii)~it exists in many stable polymorphs with varying physical properties, and hence has tunable sensitivity to various dark matter models. We show that SiC is an excellent target to search for electron, nuclear and phonon excitations from scattering of dark matter down to 10 keV in mass, as well as for absorption processes of dark matter down to 10 meV in mass. Combined with its widespread use as an alternative to silicon in other detector technologies and its availability compared to diamond, our results demonstrate that SiC holds much promise as a novel dark matter detector.
\end{abstract}

\maketitle

\section{Introduction}

The identification of the particle nature of dark matter~(DM) is one of the most pressing problems facing modern physics, and will be a key focus for high energy physics and cosmology in the coming decade~\cite{CosmicVisions,DarkMatterBRN}. In the absence of evidence for dark matter at the weak scale, interest has grown in direct searches for DM with sub-GeV mass~\cite{Essig:2011nj,Essig:2012yx,Graham:2012su,Essig:2015cda,diamonddetectors,Budnik:2017sbu,Hochberg:2016ntt,Cavoto:2017otc,Hochberg:2015pha,Hochberg:2015fth,Hochberg:2016ajh,Hochberg:2019cyy,Hochberg:2017wce,Knapen:2017ekk,Griffin:2018bjn,Schutz:2016tid,Knapen:2016cue,hertel}.

The technical challenge inherent in searching for non-relativistic, sub-GeV, weakly-interacting particles can be seen by considering the case of a classical nuclear recoil. For DM with a mass $m_{\chi}$ much smaller than the target nucleus, moving at the escape velocity of the galaxy, the maximum energy transfer for a classical elastic scattering nuclear recoil event is 
\begin{equation}
    \Delta E \approx \frac{2~\mathrm{meV}}{A_{T}}\left(\frac{m_{\chi}}{1~\mathrm{MeV}}\right)^2\,,
\end{equation}
with $A_T$ the atomic number of the target.
This motivates using lighter nuclei to increase the energy transfer, as well as new detector technologies sensitive to meV-scale energy deposits.

In Ref.~\cite{diamonddetectors}, a subset of the authors explored the ability of diamond (crystalline carbon $C$ with  $A_{T}=12$) as a detector medium to meet these criteria. The long-lived phonon states with meV energies, coupled with the light carbon nuclei, make diamond an excellent medium with which to search for dark matter. Diamond suffers from two significant drawbacks, however: it is currently difficult to produce single crystals in bulk at masses sufficient to achieve the kg-year exposures required to probe significant DM parameter space, and the non-polar nature of diamond limits the DM candidates to which it can be sensitive. 

Here we propose for the first time the use of silicon carbide (SiC) as a DM detector, as it overcomes these drawbacks.  Large wafers, and therefore also large boules, of SiC can be readily obtained at prices comparable to silicon (Si). Importantly, as a polar semiconductor, SiC has optical phonon modes which can be excited by sub-GeV DM with dark photon interactions~\cite{Knapen:2017ekk}. Furthermore, as we demonstrate in this paper, SiC behaves in most ways as a near substitute to diamond, with many relevant properties intermediate between crystalline diamond and silicon. SiC has already seen widespread adoption as a target for radiation detectors~\cite{Nava_2008}, microstrip detectors~\cite{SiCReview} and UV photodiodes~\cite{laine} as a drop-in replacement for Si in environments where greater radiation hardness, improved UV sensitivity or higher temperature operation are required. The latter two considerations are possible due to the higher band gap of SiC, $3.2$~eV, compared to 1.12~eV for Si. It is thus natural to observe the parallels between the development of Si, diamond, and SiC detector technologies, and explore the ability of future SiC detectors to search for sub-GeV DM.

Moreover, SiC is an attractive material to explore because of its polymorphism---the  large number of stable crystal structures which can be readily synthesized---and the resulting range of properties they possess. In fact SiC exhibits \textit{polytypism}---a special type of polymorphism where the crystal structures are built up from a common unit with varying connectivity between the units (see Fig.~\ref{fig:crystal_structures}). The variety of available polytypes results in a corresponding variety of physical properties relevant to DM detection, such as band gap and phonon mode frequencies. In this paper, we explore six of the most common polytypes (3C, 2H, 4H, 6H, 8H and 15R, described in detail in Section~\ref{sec:polytypes}) which span the range of variation in physical properties, and evaluate their suitability as target materials for a detector, as well as their differences in DM reach for given detector performance goals. In particular, we show that the hexagonal (H) polytypes are expected to exhibit stronger daily modulation, due to a higher degree of anisotropy in their crystal structure.

In this work we explore the potential of SiC-based single charge detectors and meV-scale microcalorimeters for DM detection. The paper is organized as follows. In Section~\ref{sec:polytypes}, we discuss the electronic and vibrational properties of the SiC polytypes explored in this work. In Section~\ref{sec:SiCDetector}, we explore the measured and modeled response of SiC crystals to nuclear and electronic energy deposits over a wide energy range, and the expected performance of SiC detectors given realistic readout schemes for charge and phonon operating modes.  Sections~\ref{sec:theoreticalFramework} and \ref{sec:results} summarize the DM models considered in this paper, and compare the reach of different SiC polymorphs into DM parameter space for nuclear recoils, direct phonon production, electron recoils and absorption processes, and also compare directional detection prospects. The high-energy theorist interested primarily in the DM reach of SiC polytypes can thus proceed directly to Section~\ref{sec:theoreticalFramework}. We find excellent DM sensitivity, comparable and complementary to other proposals, which place SiC detectors in the limelight for rapid experimental development.

\section{Electronic and Phononic Properties of SiC Polytypes}\label{sec:polytypes}

Silicon carbide is an indirect-gap semiconductor with a band gap (2.3 - 3.3 eV) intermediate between those of crystalline silicon (1.1 eV) and diamond (5.5 eV). While there exists a zincblende form of SiC, which has the same structural form of diamond and Si, there are over 200 additional stable crystal polymorphs with a range of band gap energies and  physical properties. These polymorphs broadly fall into three groups based on lattice symmetry: cubic (C), hexagonal (H), and rhombohedral (R). To compare the expected performance of these polytypes as particle detectors, we first explore how the differences in band structure between polytypes manifests in charge and phonon dynamics. 

\begin{figure}[!t]
\begin{center}
\includegraphics[width=\linewidth]{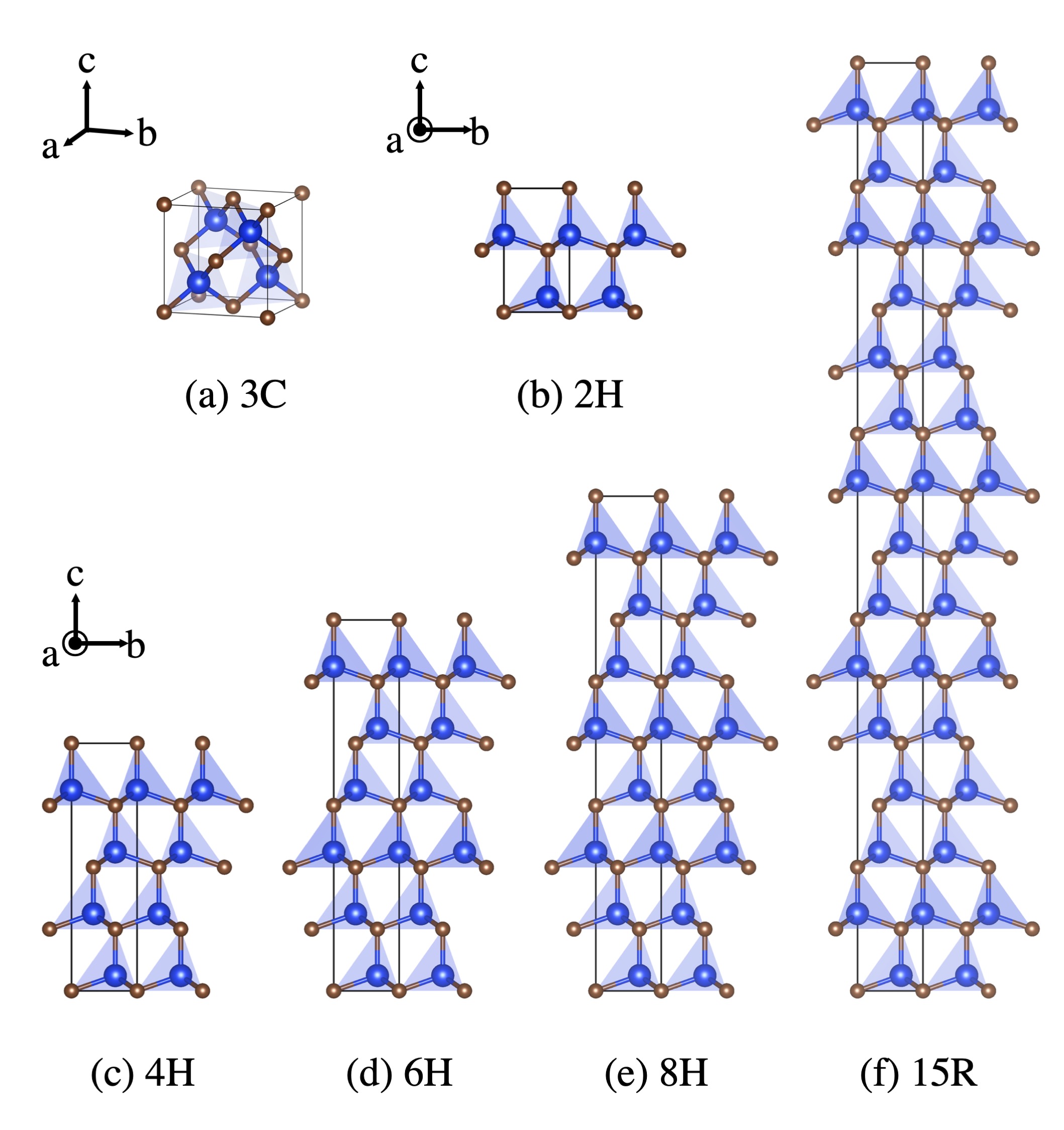}
\caption{Crystal structures of the polytypes of SiC considered in this work. Si atoms are blue and C atoms are brown.}
\label{fig:crystal_structures}
\end{center}
\end{figure}

In all SiC polytypes, the common unit is a sheet of corner-sharing tetrahedra and the polytypes are distinguished by variations in stacking sequences. The polytype 3C adopts the cubic zincblende structure with no hexagonal close-packing of the layers, whereas 2H has a wurtzite structure with hexagonal close-packing between all the layers. The different polytypes can thus be characterized by their hexagonality fraction $f_{H}$, with 2H (3C) having $f_{H}=1$ ($f_{H}=0$). This single number correlates strongly with the material's band gap, with 3C having the smallest gap, and 2H the largest gap~\cite{physprop}. The other polytypes, including those considered in this paper, consist of lattices with different sequences of hexagonal and cubic stacking layers, and can be listed in order of increasing hexagonal close-packing: 3C, 8H, 6H, 4H, 2H. The number refers the number of layers in the stacking sequence. Rhombohedral structures also occur, and these are characterized by long-range stacking order, as shown in Fig.~\ref{fig:crystal_structures}(f). Crystal structures for the polytypes considered here are shown in Fig.~\ref{fig:crystal_structures}.

The difference in stability between cubic and hexagonal stacking is very small, which can be understood as a balance between the attractive and repulsive interactions between third nearest neighbors stemming from the specific degree of charge asymmetry in the Si—C bond~\cite{Park1994}. This results in a difference in total energy between the polytypes of only a few meV per atom, therefore many crystal structures of SiC are experimentally accessible. To limit this paper to a reasonable scope, we restrict our analysis to 6 of the most common forms, as shown in Fig.~\ref{fig:crystal_structures} and with properties summarized in Table~\ref{tab:properties}. Despite the relative stability of polytypes with respect to one another, only three of these polytypes (3C, 4H and 6H) are available commercially~\cite{physprop} as of this writing; of these, 6H is the most widely available in the large wafer and crystal sizes typically employed in semiconductor processing. To capture a representative range of SiC polytype behavior in our analysis, and to observe trends in properties relevant for sub-GeV DM detection, we also include 2H, 8H, and 15R in our analysis. 

\begin{figure*}
\centering

\includegraphics[width=\textwidth]{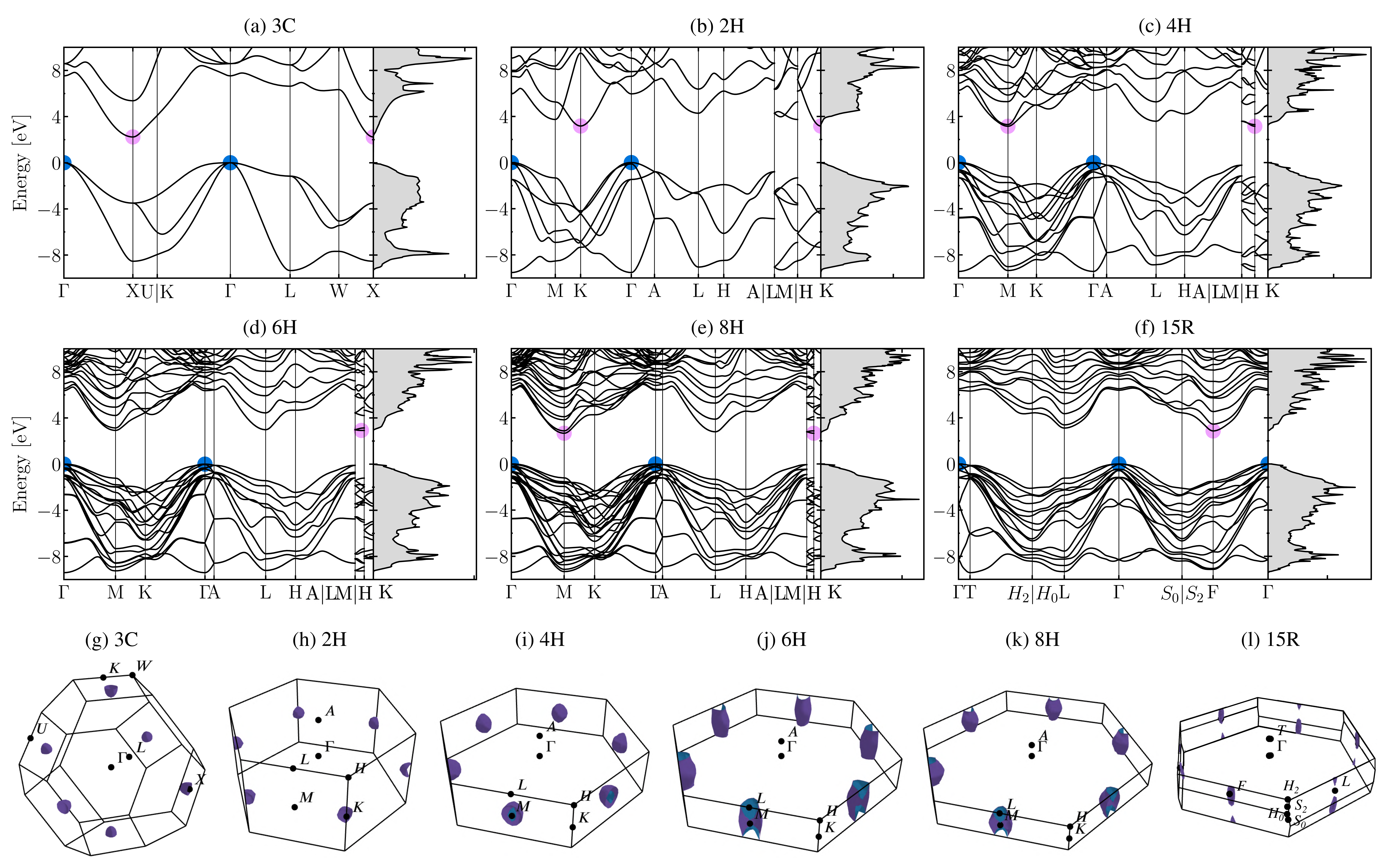}

\caption{{\bf (a) to (f):} Calculated electronic band structures of SiC polytypes, with high-symmetry paths selected using SeeK-path~\cite{HINUMA2017140}, alongside the density of states. Valence band maxima and conduction band minima are highlighted with blue and pink circles respectively. To show the conduction band valleys in momentum space, {\bf (g) to (l)} are isosurfaces of the electronic energy bands at 0.2 eV above the conduction band minima, plotted within the first Brillouin zone boundaries of the polytypes. For the positions of the high-symmetry points, see Fig.~\ref{fig:BZ}, and for details of calculations, see Appendix.~\ref{app:first_principles_calcs}.\label{fig:bandStructure}
}
\end{figure*}

\begin{figure*}[th!]
\begin{center}
\includegraphics[width=0.98\textwidth]{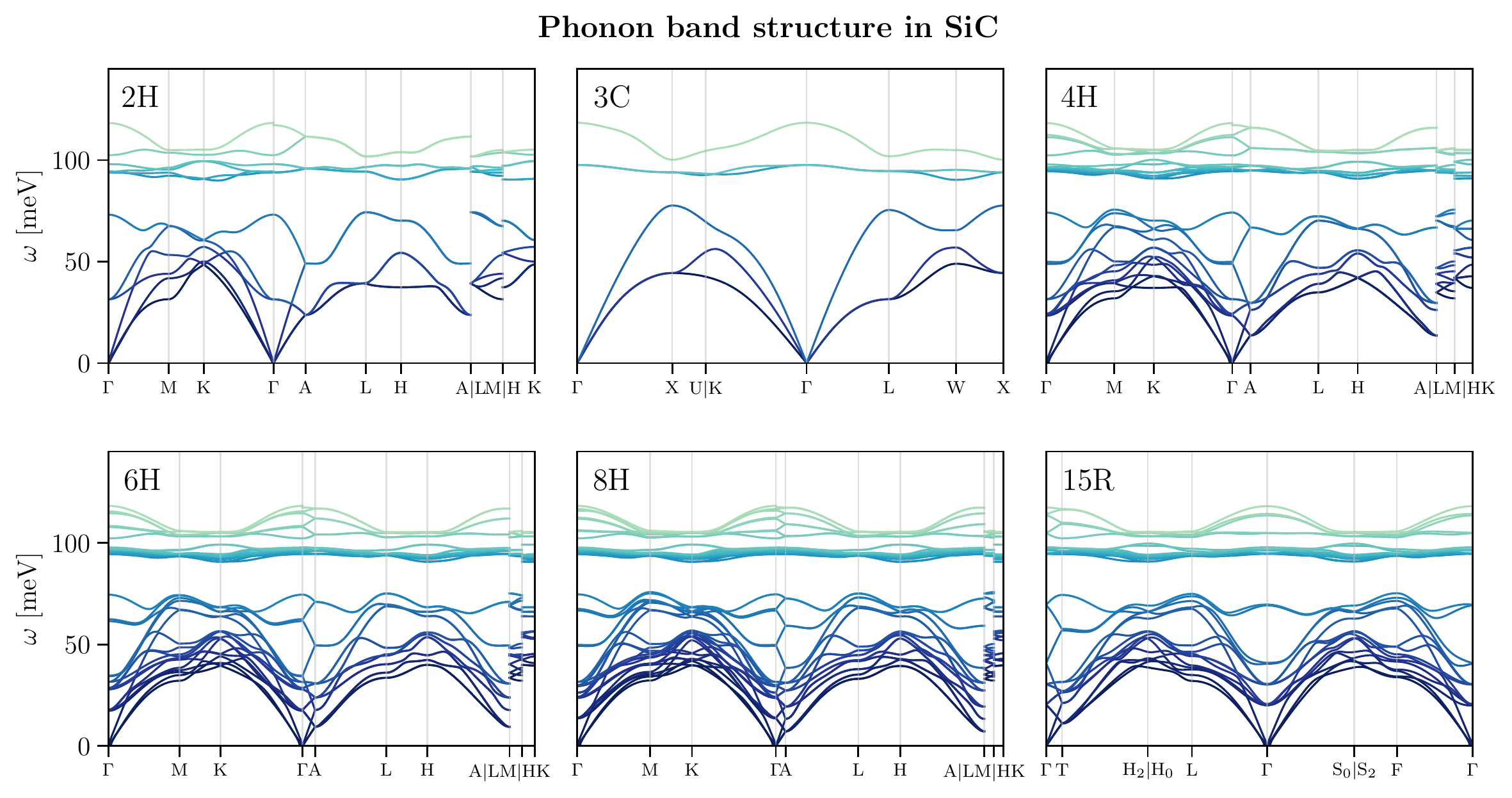}
\caption{ \label{fig:SiC_phonons} First-principles calculations of phonon band structures, with high-symmetry paths selected using SeeK-path~\cite{HINUMA2017140}. For details of calculations, see Appendix~\ref{app:first_principles_calcs}. }
\end{center}
\end{figure*}

Calculations of the interaction of various DM models with SiC requires materials-specific information for each polymorph, namely the electron and phonon spectra, to estimate sensitivity to electron and phonon interactions respectively. We calculate these quantities using state-of-the-art Density Functional Theory (DFT) calculations as described in detail in Appendix \ref{app:first_principles_calcs}. The electronic band structures for the six representative polytypes are shown in Fig.~\ref{fig:bandStructure}, and the phonon band structures are plotted in Fig.~\ref{fig:SiC_phonons}. For reference, the Brillouin zones (BZ) for the same polytypes are shown in Fig.~\ref{fig:BZ}.

The band structure of a material is important for understanding its charge or phonon dynamics, in particular charge mobility and lifetime, and phonon losses during charge propagation. As with Si, Ge, and diamond, the indirect band gap of all SiC polytypes ensures long charge lifetimes, allowing charge to be drifted and collected with a modest electric field. At low temperature, this also produces anisotropic propagation of electrons due to localized mimina in the first BZ away from the $\Gamma$ point (as shown in Si and Ge at low temperature~\cite{moffatt,StanfordSi}), which has a significant impact on charge mobility as a function of crystal orientation relative to an applied field. In Si and diamond, for example these electron valleys lie at the three X symmetry points, along the cardinal directions in momentum space. Depending on the crystal orientation relative to the electric field, spatially separated charge clusters are observed as charges settle into one of these conduction valleys.

Due to the range of stable crystal forms of SiC, in contrast to Si, diamond, and Ge, we cannot make a general statement about the location in momentum or position space of the indirect band gap in SiC (see {\it e.g.} Refs.~\cite{StanfordSi,moffatt,Moffatt:2016kok}), but we can locate the BZ minima from the band structures shown in Fig.~\ref{fig:bandStructure}. The 3C polytype, like Si and diamond, has X-valley minima, and therefore three charge valleys in the first BZ~\cite{shur2006sic}, so we can expect that the charge mobility will behave similarly to Si and diamond. The hexagonal forms, as shown in Fig.~\ref{fig:bandStructure}~(b)-(e), generally have minima along the L-M symmetry line, while the 2H polytype has minima along the K-points. All of these polytypes have 6 charge minima in the first BZ, however charge propagation in 2H will be maximally restricted to propagation along the horizontal plane of the BZ (the plane aligned with [100], [010] Miller indices). As we go to larger unit cells, charge propagation perpendicular to that plane becomes more kinematically accessible, allowing for more isotropic charge propagation. The valence bands are more consistent between polytypes, with a concentration of the valence band near the $\Gamma$ point, which is also the location of the valence band maximum for all polytypes considered. 

The dominant influence of the carbon-silicon bond (rather than the electronic orbital overlaps) on the phonon properties leads to the phonon dynamics being similar between the polytypes. Since  Si and C have near identical bonding environments, the Born effective charges are almost identical in all of the compounds considered, and so will have similar dipolar magnitudes and hence responses to dark-photon-mediated interactions. The phonon band structures for these polytypes are plotted in Fig.~\ref{fig:SiC_phonons}. While we show the entire band structure, the DM-phonon interactions are most sensitive to the phonon properties near the $\Gamma$ point. In particular, the similarities of the properties described above implies that the sensitivity of SiC for DM scattering will be similar for all polytypes. Anisotropies in the phonon band structure will give rise to differences in the directional dependence of the DM signal, as will be discussed later in this paper.

\begin{table*}[t]
\begin{tabular}{| c | c | c | c c c c c c |}
\hline
Parameter & Diamond (C) & Si & \multicolumn{6}{c|}{SiC} \\ 
\hline
Polymorph & - & - & 3C ($\beta$) & 8H & 6H ($\alpha$) & 4H & 2H & 15R \\
\hline
Crystal Structure & \multicolumn{3}{c|}{cubic} & \multicolumn{4}{c|}{hexagonal} & rhombohedral \\
 \hline
 $\rho$ (g cm$^{-3}$) & 3.51 & 2.33 & \multicolumn{6}{c|}{$\sim$3.2~\cite{SiCProperties,bertuccio}} \\
$N$ ($10^{23}$cm$^{-3}$) & 1.76 & 0.5 & \multicolumn{6}{c|}{0.96} \\
$n_e$ ($10^{23}$cm$^{-3}$) & 3.54 & 1 & \multicolumn{6}{c|}{1.95} \\
$\hbar\omega_p$ (eV) & 22 & 16.6 & \multicolumn{6}{c|}{22.1\cite{SiCPlasmon}} \\
\hline
a (c) (\AA) & 3.567 & 5.431 & 4.36 & 3.07 (20.15) & 3.08 (15.12) & 3.07 (10.05) & 3.07 (5.04) & 3.07 (37.80) \\
$f_{H}$ & 0.0 & 0.0 & 0.0 & 0.25 & 0.33 & 0.5 & 1.0 & 0.4 \\
$E_{\rm gap}$ (eV) & 5.47 & 1.12 & 2.39 & 2.7 & 3.02 & 3.26 & 3.33 & 3.0 \\
$E_{\rm gap}$ (eV)$^{[calc]}$ & & & 2.24 & 2.66 & 2.92 & 3.15 & 3.17 & 2.86 \\
$E_{eh}$ (eV) & $\sim$13 & 3.6-3.8 & 5.7 -- 7.7$^{\dagger}$ & 6.4 -- 8.7$^{\dagger}$ & 6.7~\cite{6HDet} & 7.7 -- 7.8~\cite{ivanov2005,bertuccio} & 7.8 -- 10.5 $^{\dagger}$ & 7.1 -- 9.6 $^{\dagger}$ \\
$E_{\rm defect}$ (eV) & 38--48 & 11--22 & 19 (C), 38 (Si) & & 22 (C) & 22--35~\cite{Nava_2008} & & 17--30 (C) \\
\hline
$\epsilon_{0\perp}$ & \multirow{2}{*}{5.7} & \multirow{2}{*}{11.7} & \multirow{2}{*}{9.7} &  & 9.67 & 9.76 & & \\
$\epsilon_{0\parallel}$ & & & &  & 10.03 & 10.32 & & \\
$\epsilon_{0\perp}$$^{[calc]}$ & & & \multirow{2}{*}{10.40} & 10.40 & 10.39 & 10.36 & 10.24 & 10.38 \\
$\epsilon_{0\parallel}$$^{[calc]}$ & & & & 10.80 & 10.90 & 11.06 & 11.41 & 10.96 \\
$\epsilon_{\infty\perp}$ & & & \multirow{2}{*}{6.5} &  & 6.6 & 6.6 & 6.5 & 6.5 \\
$\epsilon_{\infty\parallel}$ & & & & & 6.7 & 6.8 & 6.8 & 6.7 \\
$\epsilon_{\infty,\perp}$$^{[calc]}$ & & & \multirow{2}{*}{7.07} & 7.10 & 7.11 & 7.10 & 7.03 & 7.11 \\
$\epsilon_{\infty,\parallel}$$^{[calc]}$ & & & & 7.31 & 7.36 & 7.41 & 7.40 & 7.38 \\
$\Theta_{\rm Debye}$ (K) & 2220 & 645 & 1430 & &
1200 & 1200 & & \\
$\hbar\omega_{\rm Debye}$ (meV) & 190 & 56 & 122 & & 103 & 103 & & \\
$\hbar\omega_{\rm TO}$ (meV) & 148 & 59 & 98.7 &   & 97.7, 98.8 & 97.0, 98.8 & 95.3, 99.0 & 98.9 \\
$\hbar\omega_{\rm LO}$ (meV) & 163 & 63 & 120.5 &  & 119.7, 120.3 & 119.5, 120.0 & 120.0, 120.7 & 119.6 \\
$c_s$ (m/s) & 13360 & 5880 & 12600 & & 13300 & 13730 & & \\
$c_s$ (m/s)$^{[calc]}$ & & & 13200 & 16300 & 14300 & 14300 & 15500 & 11900 \\
$v_{d,{\rm sat}}$, $\mathrm{e^{-}}$ ($10^5$ m/s) & 2.7~\cite{bertuccio} & 1.35 & 2 & & 2 & 2 & & \\
$E_{\textrm{Bd}}$ (MV/cm) & $>$20 & 0.3 & 1.2 & & 2.4 & 2.0 & & \\ 
\hline
\end{tabular}
\caption{Bulk material properties of diamond, Si, and the SiC polymorphs considered in this work (measurements taken from Refs.~\cite{Jacoboni,kurinsky,Nava_2008,bertuccio,SiCProperties,pines} unless otherwise stated). All gaps are indirect, as discussed in the text and shown in Fig.~\ref{fig:bandStructure}. $\epsilon_{0,\infty \perp}$ ($\epsilon_{0,\infty \parallel}$) refer to relative permittivity perpendicular (parallel) to the crystal c-axis at low and high frequency, with values from Ref.~\cite{physprop}. Optical phonon energies and high-frequency permittivity are taken from Ref.~\cite{mutschke1999infrared}. $E_{eh}$ values denoted by $\dagger$ have been estimated as described in the text. Defect creation energies are from Refs.~\cite{Koike,Lucas,Barry}. Due to the differing commercial availability/utility of different polytypes, more commonly used crystal polytypes are better characterized than less common ones, and thus for the least well-studied polytypes (2H, 8H, 15R) many experimentally determined values are unavailable. Quantities denoted as [calc] were calculated in this work to fill in some of the holes in the literature.}
\label{tab:properties}
\end{table*}

Table~\ref{tab:properties} summarizes the physical properties of the polytypes shown in Figs.~\ref{fig:bandStructure} and \ref{fig:SiC_phonons} compared to Si and C. In addition, some derivative properties of the phonon band structures are summarized in the table; it can be seen that all SiC polytypes have sound speed, highest optical phonon energy and permittivity which are roughly the geometric mean of the Si and C values. These characteristics will inform our detector design and results for dark matter reach, as we now detail.

\section{Detecting Energy Deposits in SiC}\label{sec:SiCDetector}

In this section, we apply the detector performance model of Ref.~\cite{diamonddetectors} to the six representative SiC polytypes described above, and contrast expected device performance between the SiC polytypes as well as with Si, Ge and diamond targets. We begin by reviewing existing measurements and expectations for partitioning event energy into the ionization (charge) and heat (phonon) systems, relevant to reconstructing total event energy for different types of particle interactions. We then discuss expected detector performance in charge and phonon readout modes given available measurements for polytypes considered in this paper, and comment on expected performance for those polytypes without direct measurements based on band structure properties discussed above. A theorist primarily interested in the DM reach of a given SiC crystal for an assumed threshold can proceed directly to Section~\ref{sec:theoreticalFramework}.

\subsection{Particle Interactions}
\label{sec:particle_interactions}

We first turn to the expected yield for an electron recoil or nuclear recoil in SiC. As discussed in {\it e.g.} Ref.~\cite{diamonddetectors}, interactions which probe electrons or nucleons are expected to deposit differing amounts of energy in ionization and phonon systems in semiconductor detectors. This property was used by the previous generation of DM experiments to reject electron-recoil backgrounds in the search for primary nucleon-coupled WIMP DM. The resolutions in these channels required for sub-GeV DM are just now being achieved for either heat or charge in current experiments~\cite{CRESSTIII,edelweissHV,EdelweissWIMP,pd2LTD,Abramoff_2019,damicDP,DAMIC_ERDM,nucleus,strauss,Hong}, but none of these experiments can achieve the required resolutions in both channels to employ event discrimination for recoils below 1~keV in energy. For heat readout experiments, this partition is relatively unimportant, as all energy remains in the crystal and is eventually recovered as heat. For charge readout experiments, this partition is necessary to reconstruct the initial event energy, and contributes significant systematic uncertainty to background reconstruction at energies where the energy partitioning is not well-constrained. 

A convenient shorthand is to refer to the energy in the electron system as $E_e$, which is related to the total recoil energy $E_r$ according to a yield model $y(E_r)$ as $E_e = y(E_r)E_r$. As discussed in {\it e.g.} Refs.~\cite{diamonddetectors,kurinsky2020dark}, for electron recoils one has $y(E_r)=1$, while for nuclear recoils the yield is reduced due to charge shielding effects and losses to phonons and crystal defects, referred to as non-ionizing energy losses (NIEL)~\cite{LindhardDiamond}. Additionally, this yield function is actually derived with respect to the charge yield for a high-energy, minimum ionizing particle~\cite{Canali72}. These events produce a number of charge carriers $n_{eh}$ in linear proportion to event energy with the relation $n_{eh} = E_{r}/E_{eh}$, where $E_{eh}$ is taken to be a fixed property of a given material, and is the effective cost to produce a single electron-hole pair. If we define measured $E_e$ as $E_e=n_{eh}E_{eh}$, we thus see that $y(E_r)=1$ is only true, by definition, for events that obey this linear relationship.

For SiC, this factor $E_{eh}$ varies along with the band gap among the different polytypes. The charge yield from minimum ionizing particles ($\gamma$, $\beta$ and $\alpha$) in 4H SiC is explored in Ref.~\cite{Nava_2004}. The response of 3C, 4H, and 6H to lower energy X-rays is subsequently discussed in Ref.~\cite{bertuccio}. The results of both studies are consistent with a highly linear yield in electron-recoils down to ${\cal O}(10\, \rm keV)$ energies, but the pair creation energy $E_{eh}$ is only characterized for two of the polytypes, as shown in Table~\ref{tab:properties}. 

For the polytypes in which energy per electron-hole pair has not been characterized, we can predict $E_{eh}$ based on other measured properties. The generic expression for $E_{eh}$ is \cite{Klein,rothwarf,Canali72}
\begin{equation}
    E_{eh} = E_{\rm gap} + 2L \cdot \left(E_{i,e}+E_{i,h}\right) + E_{\rm ph}\,,
\end{equation}
where $L$ is a factor which depends on the dispersion curve of the conduction and valence bands, $E_{i,e}$ and $E_{i,h}$ are the ionization thresholds for electrons and holes, and $E_{\rm ph}$ are phonon losses. Ref.~\cite{Canali72} shows that, for $E_{i,e}\sim E_{i,h} \propto E_{\rm gap}$, we get the formula 
\begin{equation}
    E_{eh} = A \cdot E_{\rm gap} + E_{\rm ph}
\end{equation}
where $E_{\rm ph}$ takes on values from 0.25 to 1.2~eV, and $A$ is found to be $\sim$2.2 to 2.9. Ref.~\cite{Klein} finds, using a broader range of materials, the parameters $A\sim2.8$ and $E_{\rm ph}\sim$ 0.5 -- 1.0~eV. These allow us to predict a probable range of $E_{eh}$ values for the polytypes without existing measurements, which we summarize in Table~\ref{tab:properties}. For the detector models in this paper, we assume the values in Table~\ref{tab:properties} apply linearly for all electron-recoil events down to the band gap energy.

The response of SiC detectors to neutrons is less characterized than the electronic response. A detailed review can be found in Ref.~\cite{Nava_2008}, which we refer the reader to for more details on existing measurements. In particular, the NIEL for different particles in SiC is computed and compared to measurements for different ion beams in Ref.~\cite{Lee2003}, but this is characterized as a loss per gram, and not as a fractional energy loss compared to that lost to ionization. Ref.~\cite{dulloo2003} explores the thermal neutron response, but only a count rate is measured; a linear response with respect to fluence is measured, but there is no characterization of ionization yield on an event-by-event basis. 

We instead appeal to simulations calibrated to silicon measurements, in which the single tunable parameter with largest effect is the displacement energy threshold for freeing a nucleus from the lattice, $E_{\rm defect}$. Known and estimated values for $E_{\rm defect}$ are summarized in Table~\ref{tab:properties}. The large range is due to the difference in thresholds for the Si and C atoms; comparing the threshold values to Si and diamond, it seems that a Si atom in SiC has a diamond-like displacement threshold, while a C atom in SiC has a Si-like displacement threshold. Ref.~\cite{LindhardDiamond} calculates NIEL for Si and diamond, with the difference parameterized only in terms of defect energy. This suggests that SiC, with a defect energy intermediate between Si and diamond, will behave identically to Si and diamond above $\sim$1~keV, and give a yield below Si and above diamond for lower energy interactions.

Finally, we consider the sub-gap excitations. The most prominent features are the optical phonons, with energies of 100--120~meV, as shown in Table~\ref{tab:properties} and Fig.~\ref{fig:SiC_phonons}. An interesting property of the hexagonal polytypes is that the optical phonon energy depends on the bond direction along which they propagate, though weakly. We can expect, due to the polar nature of SiC, to see strong absorption around the optical phonon energies. We can also expect direct optical and acoustic phonon production by nuclear recoils sourced by DM interactions. 

In contrast to the large change in electron gap energy and expected pair-creation energy between polytypes, we see very little variation in phonon properties, dielectric constants, and---to some degree---displacement energy. This suggests that different polytypes will be beneficial for enhancing signal-to-noise for desired DM channels. Nucleon-coupled and phonon excitation channels would prefer higher-gap polytypes with suppressed charge production, while electron-coupled channels favor the smaller gap materials. Optimization of readout will depend on the polytype due to differences in phonon lifetime and charge diffusion length, as discussed in the next subsection, as well as differences in phonon transmission between polytypes and choice of phonon sensor. Other aspects of the design, such as capacitance and bandwidth, are constant across polytypes, somewhat simplifying the comparison of polytypes.

\subsection{Charge Readout}

The first readout mode we consider is the direct readout of charge produced in SiC crystals by low noise charge amplifiers. This mode is limited to energy deposits exceeding the gap energy of the relevant polytype, but is of interest due to the ability to run these charge detectors at higher temperatures, without requiring a dilution refrigerator, and due to the simpler readout scheme. We begin by considering the charge collection in SiC, and how the band structure of the polytypes will affect charge mobility. We then contrast the expected resolution with diamond and silicon via the resolution model of Ref.~\cite{diamonddetectors}. The resulting expected detector performance for these devices is summarized in Table~\ref{tab:detdesigns}.

\subsubsection{Charge Collection}

The primary questions for charge readout of SiC are whether complete charge collection is achievable in monolithic, insulating samples, and whether charge collection varies across polytypes. While full charge collection for the 4H polytype has been demonstrated~\cite{Vittone_2009}, detailed studies of charge collection efficiency suggest that semi-insulating samples have a fairly limited charge diffusion length at room temperature~\cite{Ruddy2008}. In Ref.~\cite{Ruddy2008} this is attributed to either recombination due to impurities or the inability to separate electron-hole pairs in the initial interaction, which causes rapid carrier recombination. More recent studies of charge collection efficiency~(CCE) in SiC radiation detectors suggest that CCE is improving with substrate quality and fabrication techniques~\cite{mandal}, though single crystal 4H-SiC still has diffusion lengths closer to polycrystalline diamond than to single crystal diamond~\cite{hodgson}. 

The only studies to demonstrate near full charge collection in SiC are Refs.~\cite{bryant,Nava_2008,bertuccio}, which all study energy deposition in thin films ($\sim$40 $\mu$m). Studies of depositions in a ten times larger detector volume in {\it e.g.} Refs.~\cite{Nava_2008,Ruddy2008} do show much reduced collection efficiency for the same bias voltage and detector readout.

These studies suggest that there remain significant bulk dislocations in these commercial wafers, which present trapping or recombination-inducing defect sites. While it is possible that charge collection will improve at lower temperatures or with higher quality substrates, there is not yet sufficient data to show this. Note that most radiation detectors to date have been constructed of 4H and 6H polytypes; it is possible that the 3C polytype, with a more symmetric band structure, could demonstrate better charge collection. A rough analogy would be comparing the charge collection of graphite to diamond, though one would expect 4H and 6H to be much more efficient than graphite. Charge collection is also likely dependent on crystal orientation relative to the BZ minima as discussed in Section~\ref{sec:polytypes}, with more efficient charge collection occurring when the electric field is aligned with an electron valley. 

For the resolution calculation presented later in this section, we will assume perfect collection efficiency; an incomplete efficiency will not affect resolution in the single-charge limit, but will instead reduce effective exposure. To minimize the effect of limited charge collection on detector performance, we require the drift length (detector thickness) to be equal to or less than the diffusion length of the target charge carrier at the design voltage.

To make this more quantitative, one can model the CCE in terms of a few measured parameters. Given a carrier mobility $\mu$ (in principle different for electrons and holes) and saturation velocity $v_{d,{\rm sat}}$, we use an ansatz for carrier velocity as a function of voltage $V$:
\begin{equation}
    v_d(V,d) = \left[\frac{1}{v_{d,{\rm sat}}}+\frac{d}{\mu V}\right]^{-1}
\end{equation}
where $d$ is the detector thickness. This gives the drift length $D = v_d\tau_{\rm scat} \rightarrow v_{d,{\rm sat}}\tau_{\rm scat}$ in the high-field limit~\cite{Nava_2008}, where $\tau_{\rm scat}$ is the carrier scattering lifetime. Given this drift length, we can model the CCE as~\cite{bryant}
\begin{equation}
    {\rm CCE} = \frac{D}{d}\left[1-\exp\left(-\frac{d}{D}\right)\right]
\end{equation}
where for long diffusion length ($D \gg d$) we have CCE$\sim1$. For short diffusion length, and in the small-field limit, we find that charge collection goes as
\begin{equation}
    {\rm CCE} \approx \frac{\mu V \tau_{\rm scat}}{d^2} = \frac{\mu \tau_{\rm scat}}{d}E \ll 1
\end{equation}
with $E$ the electric field in the bulk. This tells us that when the gain is linear in voltage, the inferred CCE will be small and the effective diffusion length is much shorter than the crystal thickness. 

The best measure of the drift constant $\mu\tau_{\rm scat}$ in 4H-SiC (the only polytype for which detailed studies are available) was found to be $\mu\tau_{\rm scat}\sim 3\times 10^{-4}$ cm$^2$/V, and for a saturation drift field of 8~kV/cm, we find a maximum drift length $D\sim 2.4$~cm~\cite{bryant}. While this does imply full charge collection for devices up to 1~cm thick, the very high voltages required are likely to induce some measure of charge breakdown, despite the very high dielectric strength of SiC.  The devices studied in Refs.~\cite{bryant,Nava_2008,bertuccio} are all thin films which did not break down at field strengths in this regime; however, for low temperature operation of these devices, voltages of this magnitude are atypical for monolithic, gram-scale detectors. Ref.~\cite{bertuccio} suggests there is a very small difference in mobility between the 3C, 4H, and 6H polytypes, but it is possible that the more isolated valleys of 3C, and different growth process, may lead to larger charge lifetime. To better determine the polytype best suited to charge collection, more studies of drift length in high-purity samples are needed.

\subsubsection{Charge Resolution}

\begin{table*}[t]
    \centering
    \begin{tabular}{|l|c|c|c|c|c|c|}
    \hline
    Readout & Design & Dimensions & Mass (g) & Temp. (K) & $V_{\rm bias}$ & $\sigma_{q}$ \\
    \hline
    \multirow{4}{*}{Charge} & Single Cell & $1.0~{\rm cm~side~length} \times 0.5~{\rm cm~thick}$ & 1.6 & \multirow{4}{*}{4.2~K} & 4~kV & 1.4$e^{-}$ \\
    & Single Cell & $0.5~{\rm cm~side~length} \times 0.5~{\rm cm~thick}$ & 0.4 & & 4~kV & 0.5$e^{-}$ \\
    & Single Cell & $1.0~{\rm cm~diameter} \times 1.5~{\rm cm~thick}$ & 4.8 & & 500~V & 0.5$e^{-}$ \\
    & Segmented & $0.2~{\rm cm~side~length}\times 0.2~{\rm cm~thick}$ & 0.025 & & 50~V & 0.25$e^{-}$/segment \\
    \hline
    \end{tabular}
    \caption{Summary of the detector designs discussion for charge readout. Voltage bias for the charge designs should be high enough to ensure full charge collection. For the lower two charge readout designs, improved charge lifetime is assumed, allowing for lower voltage bias and thicker crystals. 
    We note that, due to the relatively high dielectric constant of SiC, the optimal geometry (given current readout constraints) is such that cells have a thickness greater than or equal to the side length in order to minimize capacitance per unit mass. }
    \label{tab:detdesigns}
\end{table*}

Recalling the model for charge resolution from Ref.~\cite{diamonddetectors}, the minimum resolution of a charge integrating readout is completely determined by the noise properties of the amplifier, the bias circuit, and the capacitance of the detector ($C_{\rm det}$) and amplifier ($C_{\rm in}$) (see {\it e.g.} Ref.~\cite{shuttThesis}):
\begin{equation}
\sigma_{q} \ge \frac{N_{v}(C_{\rm det}+C_{\rm in})}{\epsilon_q \sqrt{\tau}},
\end{equation}
where $N_{v}$ is assumed to be a flat voltage noise spectral density of the amplifier in $V/\sqrt{\rm Hz}$, $\epsilon_q$ is the CCE and $\tau$ is the response time of the detector and readout. For an integrator, the readout time $\tau$ is determined by the rate at which the input is drained by some bias resistor $R_b$, and thus $\tau = R_b (C_{\rm det}+C_{\rm in})$. 

Following the discussion of Ref.~\cite{diamonddetectors}, the current best cryogenic high electron mobility transistor (HEMT) amplifiers \cite{Phipps} allow for a detector resolution of
\begin{equation}
\sigma_{q}\approx (28\;\mathrm{e^{-}h^{+}\;pairs})\left(C_{\rm det}/(100~\mathrm{pF})\right)^{3/4}\,,
\end{equation}
where we have enforced the optimal design condition $C_{\rm in} = C_{\rm det}$, and we assume full CCE can be achieved (this is ensured by limiting thickness to 1~cm). Note that if this resolution for 100\% CCE is sub-electron, it affects the effective resolution on the input signal rather than the resolution of the readout.\footnote{For the case of incomplete charge collection for detectors with single electron resolution, the resolution is not smeared due to Poisson fluctuations, but the conversion from charge to energy scale requires folding in charge collection statistics. For detectors without single electron resolution, limited CCE effectively contributes an additional Poisson smearing to the Gaussian noise PDF.}

We give example design parameters for a charge detector in Table~\ref{tab:detdesigns}. We consider both monolithic and segmented detectors, the latter necessary to achieve statistically significant sub-electron resolution at reasonable detector mass. One benefit of SiC over {\it e.g.} diamond is that larger crystals are readily available commercially, allowing for designs with ${\cal O}$(1e$^{-}$) resolution. All designs are limited to 1.5~cm thickness---the largest thickness currently available, and to ensure full charge collection at a field of 8~kV/cm. We likewise assume significant voltage bias in our designs for this reason, and in our latter charge designs assume improvements can be made in drift length by improving mean carrier lifetime through advances in crystal growth technology (see Ref.~\cite{bryant} for a more detailed discussion). 

The size and resolution of the segmented design suggest that development of SiC detectors with a skipper CCD readout is likely a more straightforward development path; large-scale fabrication of SiC devices has been available for decades, and feature sizes required are fairly modest. Such developments would be useful for employing large-area SiC sensors as beam monitors and UV photon detectors, and would be complementary to Si substrates for dark matter detection thanks to the reduced leakage current due to the higher gap and dielectric strength of SiC.

\subsection{SiC Calorimetry}

The most promising direction for application of SiC to dark matter searches is direct phonon readout at cryogenic temperatures. The intrinsic phonon resolution $\sigma_{\rm ph}$ is the primary metric for determining DM reach in this case. Here we take a technology-agnostic approach to computing expected phonon resolution; rather than calculating resolutions for a specific detector technology, we will relate resolution to phonon collection efficiency, intrinsic phonon properties of the material, and input-referred noise equivalent power (NEP) of the readout. For the last quantity, we will use reference values comparable to those currently achievable by a range of cryogenic sensing techniques. We will compare this with currently achieved resolutions using other crystals, as well as technologies of sufficiently low noise temperature to achieve sub-eV resolutions, and suggest form factors and noise temperature targets for the various thresholds discussed for DM sensitivities later in this paper. 

Following this parameterization, we thus calculate resolution as
\begin{equation}
    \sigma_{\rm ph} = \frac{1}{\epsilon_{\rm ph}}\sqrt{S_{\rm ph}\tau_{\rm pulse}}
\end{equation}
where $\epsilon_{\rm ph}$ is the energy efficiency for phonon collection, $S_{\rm ph}$ is the NEP for the readout in $W^2/{\rm Hz} \propto {\rm eV}^2/{\rm s}$, and $\tau_{\rm pulse}$ is the duration of the signal in seconds.\footnote{$\tau_{\rm pulse}$ can also be thought of as the inverse of the bandwidth ($\tau_{\rm pulse}=2\pi/\omega_{\rm pulse}$). We use $\tau_{\rm pulse}$ rather than $\omega_{\rm pulse}$ for easier comparison with sensor response time $\tau_{\rm sensor}$, given that $\tau_{\rm pulse} = \tau_{\rm ph}+\tau_{\rm sensor}$, where $\tau_{\rm ph}$ is the phonon signal time.} This is similar to the detector treatment in Refs.~\cite{Hochberg:2015fth,diamonddetectors}, and uses the same terminology as for Transition Edge Sensor~(TES) noise modeling~\cite{Irwin}, but is written more generally for ease of comparison between readout technologies. 

\subsubsection{Phonon Collection Efficiency}

The primary metric which determines whether a material will allow for efficient phonon collection is the phonon lifetime $\tau_{\rm life}$. As discussed in Ref.~\cite{diamonddetectors}, for pure crystals at low temperature, the lifetime is limited primarily by boundary scattering.  This scaling for the phonon lifetime can be inferred from thermal conductance data, given knowledge of material density, sound speed and crystal size. A model for the thermal conductance and its relation to the phonon lifetime is described in Appendix~\ref{app:kappa}. 

For diamond, it was found that boundary scattering is dominant for phonons of $\leq$10~K, implying that the bulk mean free path for phonons at and below this energy ($\lesssim 1~$meV) is much longer than the typical crystal length scale (1-10 mm) \cite{diamonddetectors}. For SiC, the thermal conductivity (at least that of 6H~\cite{SLACK1973321}) and sound speeds are close to that of diamond, so we can infer that SiC will similarly be limited by boundary scattering, at least for phonons near the pair-breaking energy of the superconducting phonon sensors. The phonon band structure calculations from Section~\ref{sec:polytypes} (calculation details in Appendix~\ref{app:first_principles_calcs}) were used to verify that low-energy acoustic phonons (below 2 THz) have bulk lifetimes much longer than their collection timescales ($>10$~ms). For 3C, the calculated average phonon lifetime within 0-2~THz is of the order 30~ms at 2~K. In the hexagonal polytypes, the phonon lifetimes will be smaller because of increased scattering from the variations in stacking sequences inherent to the structures, but initial calculation results at 10~K indicate that the 2H lifetimes will be within an order of magnitude of those for 3C.

Assume a detector in the form of a prism of thickness $\eta$ and area $A$. With only one type of phonon absorber, the phonon collection time-constant is~\cite{Hochberg:2015fth,diamonddetectors}
\begin{equation}
    \tau_{\rm collect} = \frac{4\eta}{f_{\rm abs}\bar{n}_{\rm abs}c_s}
\end{equation}
where $f_{\rm abs}$ is the fraction of the detector surface area covered by phonon absorber material and $\bar{n}_{\rm abs}$ is the transmission probability between the bulk and the absorber. $\bar{n}_{\rm abs}$ is calculated in detail in Appendix~\ref{app:ptrans} with values in Table~\ref{tab:trans}. As a basis for comparison, the worst-case scenario that phonons are completely thermalized at the crystal sidewalls gives a bound on the phonon lifetime of $\tau_{\rm life}\gtrsim \eta/c_s$, a single phonon crossing time across the crystal. In the following we will explore the case where boundaries are highly reflective, in which case $\tau_{\rm life} \gg \tau_{\rm collect}$, as well as the case where boundaries are sources of phonon losses, in which $\tau_{\rm collect} \gtrsim \tau_{\rm life}$.

In all cases, the phonon pulse time is determined by combining phonon collection time with phonon lifetime as~\cite{Hochberg:2015fth}:
\begin{equation}
    \tau_{\rm pulse}^{-1} \approx \tau_{\rm ph}^{-1} = \tau_{\rm life}^{-1} + \tau_{\rm collect}^{-1}\,,
\end{equation}
where we assume the sensor is much faster than the timescale of phonon dynamics ($\tau_{\rm ph} \gg \tau_{\rm sensor}$). Then the overall collection efficiency is then
\begin{equation}
    f_{\rm collect} = \frac{\tau_{\rm pulse}}{\tau_{\rm collect}} = \frac{\tau_{\rm life}}{\tau_{\rm life}+\tau_{\rm collect}}\,.
\end{equation}
The total detector efficiency is then given as a product of the conversion and readout efficiencies,
\begin{equation}
    \epsilon_{\rm ph} = f_{\rm collect}\epsilon_{qp}\epsilon_{\rm trap}
\end{equation}
where $\epsilon_{qp}$ is the efficiency of generating quasiparticles in the phonon absorber, and $\epsilon_{\rm trap}$ is the efficiency of reading out these quasiparticles before they recombine. 
$\epsilon_{qp}$ has a generic limit of 60\% due to thermal phonon losses back into the substrate during the quasiparticle down-conversion process~\cite{Guruswamy_2014}, though it rises to unity as the captured energy approaches $2\Delta\sim 7k_bT_c/2$, the Cooper pair binding energy for an absorber at $T_c$. Meanwhile, $\epsilon_{\rm trap}$ is technology dependent. For quasiparticle-trap assisted TESs, $\epsilon_{\rm trap}$ is limited by quasiparticle diffusion and losses into the substrate, while for superconducting resonators such as KIDs, $\epsilon_{\rm trap}$  is governed by the response time of the resonator compared to the recombination lifetime of the quasiparticles. 

\subsubsection{Material-Limited Resolution}

Since different readout technologies are possible, here we instead focus on the material-limited resolution of a SiC detector. We will thus consider an idealized phonon readout with a response time much faster than the characteristic phonon timescale and a benchmark noise temperature near that currently achieved by infrared photon detectors and prototype TES calorimeters. Taking a single sensor with NEP $\sqrt{S_{s}}\sim 10^{-19}$~W/$\sqrt{\mathrm{Hz}}$ \cite{lowNEP,THzSinglePhoton,fink2020characterizing},\footnote{Here we are scaling the noise power measured in the reference to the effective volume of a single QET as characterized in Ref.~\cite{Hong}.} and assuming our idealized readout is limited by the timescale of phonon dynamics, we find a single-sensor resolution of
\begin{align}
    \sigma_{\rm ph} &\approx 10^{-19}~\mathrm{W}/\sqrt{\mathrm{Hz}} \frac{1}{\epsilon_{\rm ph}}\sqrt{\tau_{\rm pulse}} \\
    &\approx 10~\mathrm{meV}\frac{1}{f_{\rm collect}\epsilon_{\rm trap}}\sqrt{\frac{\tau_{\rm pulse}}{100\; \mathrm{\mu s}}}\, \\
    & \approx \frac{10~\mathrm{meV} }{\epsilon_{\rm trap}} \sqrt{ \frac{\tau_{\rm collect}^2}{\tau_{\rm pulse} \times 100~\mathrm{\mu s} } }
\end{align}
where we have set $\epsilon_{\rm qp} = 0.6$.
We thus see that the challenges for excellent resolution are to achieve high internal quantum efficiency between phonon absorber and phonon sensor ($\epsilon_{\rm trap}$), and to ensure fast phonon collection (short $\tau_{\rm collect}$, with $\tau_{\rm life}$ not too small compared to $\tau_{\rm collect}$).

\begin{table*}[]
    \centering
    \begin{tabular}{|c|c|c|c|c|c|}
    \hline
       & & \multicolumn{4}{c|}{Design}  \\
       \hline
        & Parameter & A & B & C & D \\
        \hline
        & Polytype & 6H or 4H & 3C & 3C & Any \\
        \hline
        & Phonon Absorber & \multicolumn{2}{c|}{Al} & \multicolumn{2}{c|}{AlMn} \\
        $2\Delta$ & Pair-Breaking Threshold & \multicolumn{2}{c|}{700~$\mu$eV} & \multicolumn{2}{c|}{60~$\mu$eV} \\
        \hline
        $\epsilon_{\rm qp}$ &  Efficiency to generate quasiparticle in absorber & \multicolumn{4}{c|}{60\%} \\
        $\epsilon_{\rm trap}$ & Efficiency to readout quasiparticle in absorber & \multicolumn{4}{c|}{75\%} \\
        $\tau_{\rm ac}$ & Acoustic Phonon Lifetime (crystal limited) & \multicolumn{4}{c|}{$>30$~\us} \\
        \hline
        $\tau_{\rm life}$ &  Assumed phonon lifetime (boundary limited) & \multicolumn{2}{c|}{$\sim$100~\us} & \multicolumn{2}{c|}{$\sim$1~\us} \\
        $\sqrt{S_{s}/A_s}$ & Noise power per unit sensor area  ($\rm W/{mm \cdot Hz^{1/2}}$)  & \multicolumn{2}{c|}{$10^{-19}$} & \multicolumn{2}{c|}{$10^{-20}$\footnote{This noise power is the best currently achievable in any quantum sensor; see for example Refs.~\cite{lowNEP,THzSinglePhoton}.}} \\
        $\sqrt{S_{s}}$  & Noise power per sensor ($\rm meV/s^{1/2}$) & \multicolumn{2}{c|}{600} & \multicolumn{2}{c|}{60} \\
        \hline
        $A$ & Detector area & 45 $\rm cm^2$ & 5 $\rm cm^2$ & \multicolumn{2}{c|}{1~$\rm cm^2$} \\
        $\eta$ & Detector thickness & 1~cm & 1~cm & \multicolumn{2}{c|}{4~mm}  \\
        $\bar{n}_{\rm abs}$ & Transmission probability to absorber & 0.83 & 0.94 & \multicolumn{2}{c|}{$\sim$0.94\footnote{AlMn films are primarily Al, containing $<$1\% Mn~\cite{deiker}, and we assume the transmission coefficient will be approximately equal to the pure Al case.}} \\
        $f_{\rm abs}$ & Fractional coverage of detector surface with absorber & 0.1 & 0.7 & \multicolumn{2}{c|}{0.95} \\
        $N_{s}$  & Number of sensors & 450 & 350 & \multicolumn{2}{c|}{95} \\
        $\tau_{\rm collect}$ & Time scale to collect ballistic phonons & 34~\us & 4.3~\us & \multicolumn{2}{c|}{1.3~\us} \\
        $\tau_{\rm pulse}$ & Time scale of phonon pulse & 25~\us & 4.2~\us & \multicolumn{2}{c|}{0.5~\us} \\
        $f_{\rm collect}$ & Collection efficiency into absorber & 74\% & 95\% & \multicolumn{2}{c|}{45\%} \\
        $\epsilon_{\rm ph}$ & Total signal efficiency for detector & $\sim$30\% & $\sim$40\% & \multicolumn{2}{c|}{20\%} \\
        \hline
        & Detector mass & 145~g & 16~g & \multicolumn{2}{c|}{1~g} \\
        \hline
        $\sigma_{\rm ph}$ & Resolution on phonon signal & 200~meV & 50~meV & 2~meV & $\sim$0.5~meV\footnote{This assumes 5 or fewer sensors can be used to read out the total phonon signal, and that the phonon dynamics are still the bandwidth limiting timescale.} \\
        \hline
    \end{tabular}
    \caption{Reference phonon detector designs. Designs A and B assume performance parameters for currently demonstrated technology, while design C assumes an improvement by a factor of 10 in noise equivalent power per sensor. The main limitation affecting design C is that, for very low thresholds, effective phonon lifetime may be as short as a few times the crystal crossing time, due primarily to phonon thermalization at crystal boundaries. If significant phonon thermalization is allowed to occur, the phonon resolution will quickly be limited by statistical fluctuations in signal collection efficiency rather than sensor input noise. For this reason, our third design assumes an effective phonon lifetime equivalent to 3 crystal crossings, a lower gap absorber, and 95\% sensor coverage. The limited absorption probability and realistic constraints on sensor area coverage severely limit the overall efficiency ($\epsilon_{\rm ph}$) of the design relative to the other two reference designs. The polytype selection is primarily determined by the impedance match between the substrate and phonon absorber. All polytypes are fairly well matched to the chosen absorbers, but the 3C polytype is a close match and maximizes phonons absorbed per surface reflection.}
    \label{tab:calorimeterDesigns}
\end{table*}

Realistically, readout noise power scales with sensor volume, and we can tie the above benchmark noise temperature to a reasonable sensor area. For current technology, a single superconducting sensor can typically readout about $A_{s}\sim$1~mm$^2$ of area, and thus we can parameterize the above equations more accurately in terms of this sensor area and detector geometry. We find that 
\begin{align}
    f_{\rm abs} &= N_{s}A_{s}/A \\
    S_{\rm ph} & =N_{s}S_{s} = \frac{f_{\rm abs}A}{A_s}S_{s}.
\end{align}
For the above reference noise temperature and assuming $\tau_{\rm life} \gg \tau_{\rm collect}$, this gives an energy resolution of
\begin{align}
    \sigma_{\rm ph}
    &\approx \frac{6~\mathrm{meV}}{\epsilon_{\rm trap}}\sqrt{\frac{V}{100~\mathrm{mm^3}}\frac{1~\mathrm{mm^2}}{A_s}\frac{0.95}{\bar{n}_{\rm abs}}\frac{14 \, \mathrm{km/s}}{c_s}} \label{eq:reslonglife}
\end{align}
where $V$ is the detector volume. This is the generic result that an ideal athermal detector has a resolution that scales as $\sqrt{V}$ for a given readout technology, and as $(\bar{n}_{\rm abs}c_s)^{-1/2}$ for a given crystal/phonon absorber coupling.

In the opposite limit where $\tau_{\rm life} \ll \tau_{\rm collect}$, we find the resolution scales as 
\begin{align}
      \sigma_{\rm ph} \approx \frac{13 ~\mathrm{meV}}{\epsilon_{\rm trap}}\left(\frac{0.95}{\bar{n}_{\rm abs}}\right)\sqrt{\frac{1}{f_{\rm abs}} \frac{V}{100~\mathrm{mm^3}} \frac{(\eta/c_s)}{\tau_{\rm life}} } 
    \label{eq:resshortlife}
\end{align}
where we have again used $A_s = 1~{\rm mm}^3$ and the sound speed in SiC. In this case, the detector design relies on high surface coverage $f_{\rm abs}$ to maximize phonon collection, and the resolution is more sensitive to the phonon transmission probability, $\bar n_{\rm abs}$. For the chosen parameters this is only about twice the resolution of the long-lived phonon case, but it is more sensitive to details of sensor coverage and will be more sensitive to phonon losses both in the crystal and at the crystal-absorber interface.

These estimates assume that the detector in question can be read out with sub-microsecond precision (such that $\tau_{\rm sensor} \ll \tau_{\rm ph}$, as stated earlier), while sensors at this level of power sensitivity are not necessary capable of being read out at this rate \cite{kurinskyThesis,fink2020characterizing}; we comment on this more below. Finally, this term does not include phonon shot noise, which will be a significant source of additional variance in the limit of small phonon lifetime. All of this goes to say that the ideal detector design will be highly dependent on whether phonons are completely thermalized at the boundaries, or if there is a reasonable chance of reflection of athermal phonons such that there's a non-zero survival probability for each surface interaction.

\begin{figure*}[th!]
    \centering
    \includegraphics[width=0.47\textwidth]{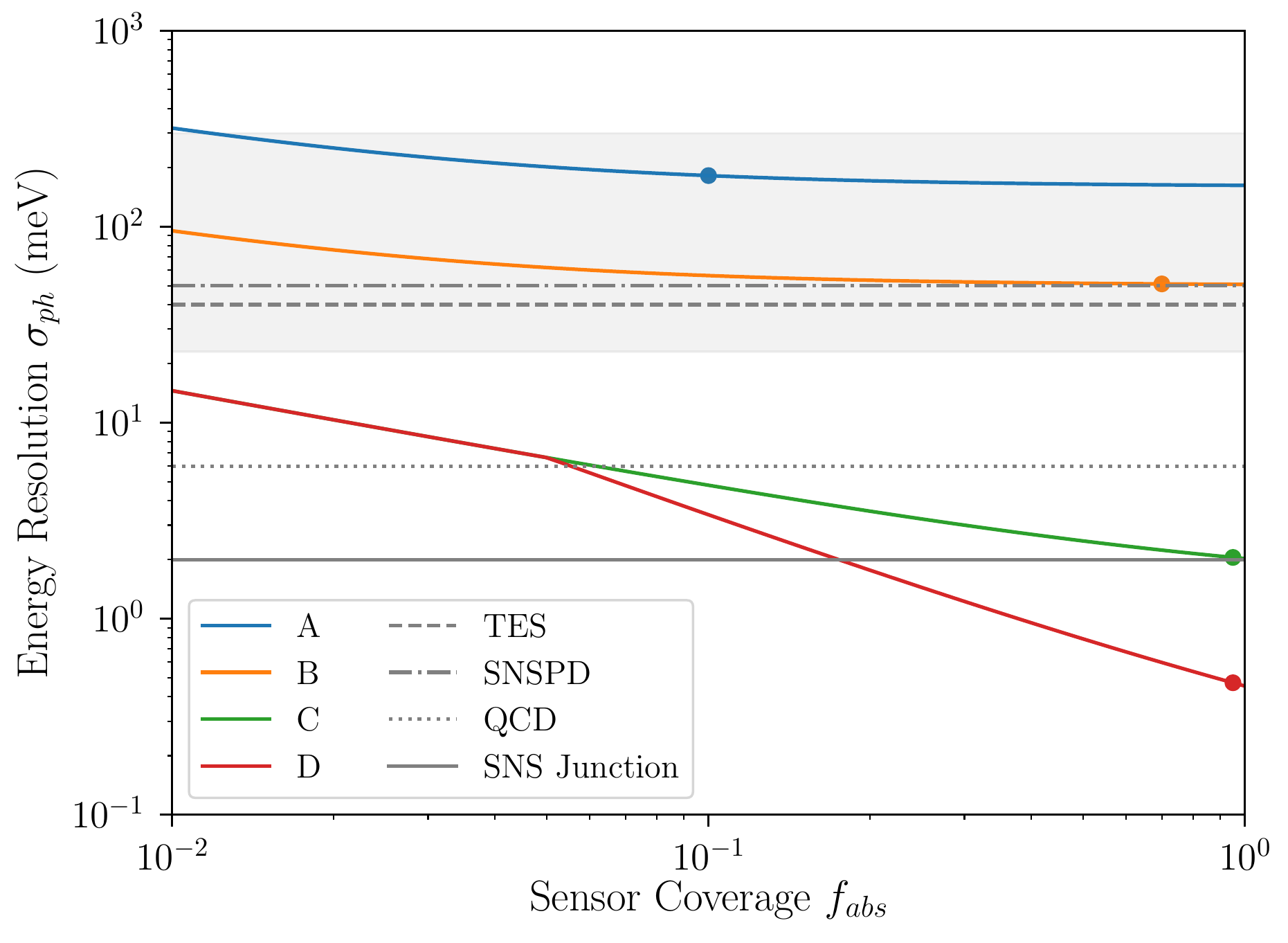}
    \includegraphics[width=0.45\textwidth]{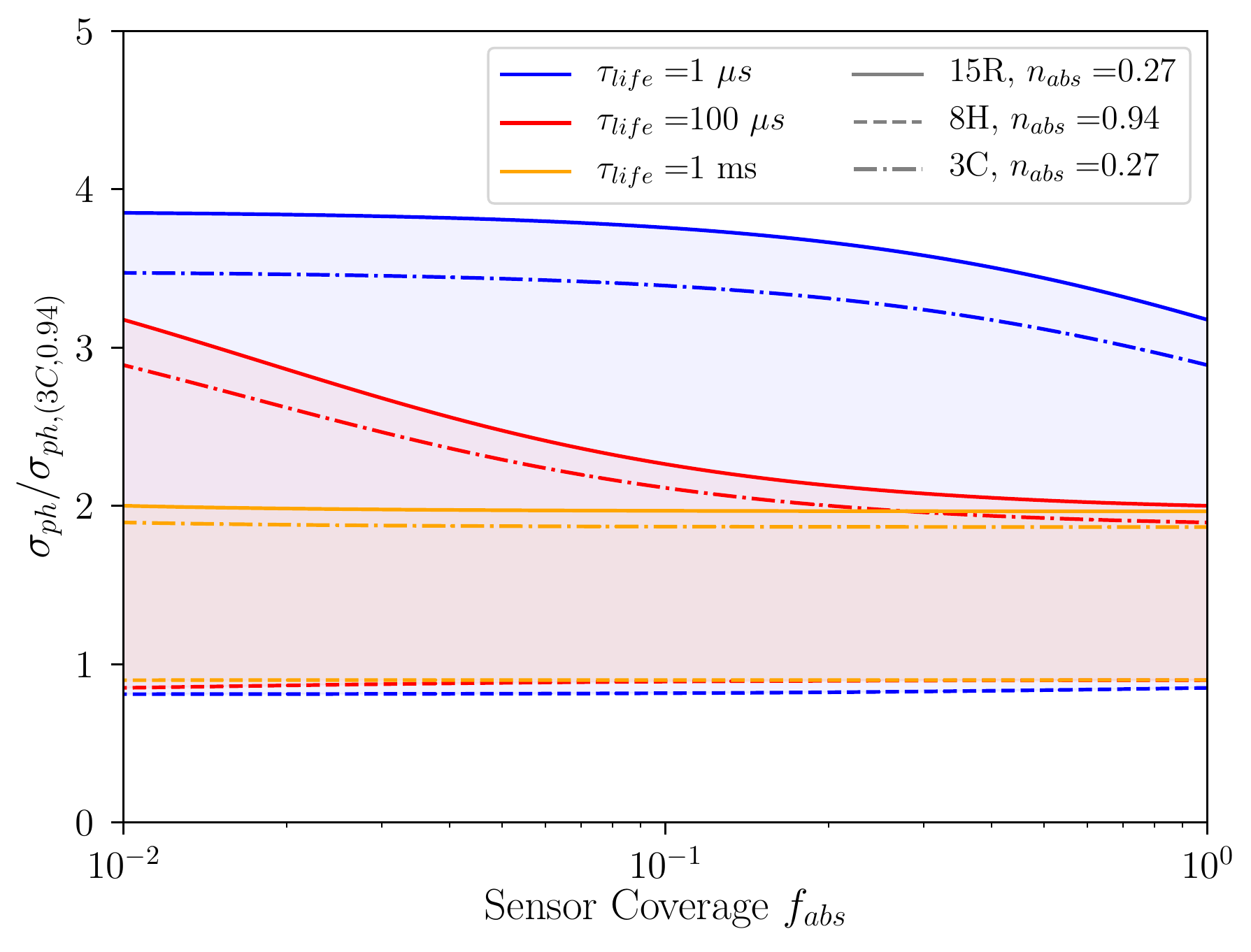}
    \caption{{\bf Left:} The dependence of energy resolution on fractional surface area covered by sensors ($f_{\rm abs}$) is shown for each of the designs from Table~\ref{tab:calorimeterDesigns}. Dots indicate the resolutions quoted in the table. For designs C and D, the resolution is the same in the small $f_{\rm abs}$ limit, or where there are 5 or fewer sensors; in this limit, all sensors are assumed necessary to reconstruct events. Meanwhile, for larger $f_{\rm abs}$, the improved scaling of design D over design C results from the fixed number of sensors read out (here taken to be 5) as the detector bandwidth is increased. Also shown are devices with the current best demonstrated noise power and resolution. The TES and SNSPD benchmarks come from Refs.~\cite{fink2020characterizing,Hochberg:2019cyy}, where the shaded band corresponds to the detectors listed in Ref.~\cite{fink2020characterizing}, and the lines correspond to best DM detector performance from the respective references. We also include two superconducting photon detectors optimized for high detection efficiency for THz photons, where the best demonstrated NEP, roughly $10^{-20}$W/$\rm \sqrt{Hz}$ in both cases, is comparable to the NEP assumed for designs C and D. The quantum capacitance detector (QCD) is from Ref.~\cite{THzSinglePhoton}, and the SNS junction is from Ref.~\cite{lowNEP}. {\bf Right:} The relative change in resolution as polytype and interface transmission are changed for a range of (surface-limited) phonon lifetimes, compared to the nominal, impedance-matched 3C/Al design at the chosen phonon lifetime. The resolutions of these devices are best case scenarios for perfect phonon detection efficiency, and thus represent a lower resolution limit for the given technology.}
    \label{fig:energyRes}
\end{figure*}

Table~\ref{tab:calorimeterDesigns} summarizes our four reference designs, with resolutions varying from 200~meV (design A) down to 500~$\mu$eV (design D). Designs A and B assume the device is read out by phonon sensors comparable to those that have currently been demonstrated, and the resolution is in the long phonon lifetime regime of Eq.~\eqref{eq:reslonglife}. The design thresholds for these devices assume that the majority of initial phonons lie far above the absorption gap of the phonon sensor ($2\Delta \sim$~0.7~meV in Al) and that down-conversion at the crystal surfaces has a small impact on total phonon energy absorbed by the sensors.  The resolution scaling between A and B then comes just from relative reduction of crystal volume.

For designs C and D, we consider an initial phonon energy small enough that only a few phonon scattering events can occur before phonons are absorbed. This implies we will be in the short lifetime regime and need to have large coverage $f_{\rm abs} \to 1$ to avoid substantial signal loss. To attain resolutions low enough to observe single phonon production, we also assume here an order of magnitude decrease in noise power over currently demonstrated phonon sensors. Design C obeys the scaling of Eq.~\eqref{eq:resshortlife}. Design D has the same detector geometry, but here we assume that only 5 or fewer sensors need to be read out to reconstruct a signal. This provides an improvement in resolution by reducing the number of sensors read out by a factor of 20, without necessarily changing the detector or sensor properties. The timescale for this process is still the phonon crossing time for the crystal; additional resolution reduction could still be accomplished by reducing the size of the crystal, though gains would be fairly modest. 

In addition, the resolution for sensors using quasi-particle traps to read out phonons will hit a floor at the pair-breaking energy of the phonon absorber, which for Al is $2\Delta \sim0.7$~meV, and for AlMn (with a $T_c$ around 100 mK~\cite{deiker}) is $\sim0.06$~meV (see also Table~\ref{tab:calorimeterDesigns}). For this reason we assume that detectors with resolution $<$~50~meV (designs C and D) will need to transition to lower-gap materials; this ensures that the phonons they intend to detect can break $\gtrsim$~100 quasiparticles per sensor to minimize shot noise contributions to the noise budget. 

In Fig.~\ref{fig:energyRes}, we show the scaling of resolution with sensor area, along with our reference designs, in comparison to currently achieved resolutions by an array of superconducting sensors. These scalings are based on a fixed sensor form factor, with the given noise performance corresponding to an areal coverage of $A_s\sim 1 {\rm mm}^2$, and the lines assume a fixed power noise per unit area (as described earlier in this section) for a variety of sensor coverage and crystal form factors. In all cases, significant enhancements in sensor noise power are required to achieve less than 100~meV resolutions even for gram-scale detectors, and we note that the detection thresholds for these detectors will be a multiple of these resolutions. This limitation is not specific to SiC but broadly applies to any solid-state phonon calorimeters using superconducting readout. In particular, we note that only designs C and D would be expected to detect single optical phonon excitations. 

The choice of different polytypes in the detector designs in Table~\ref{tab:calorimeterDesigns} lead to minor changes in expected resolution due to sound speed (which varies by 25\% between polytypes) and impedance matching between the crystal and the absorber (a difference of less than 20\%). The same sensor design ported to different polytypes can therefore vary by up to around a factor of 2 in resolution, a non-trivial amount but small compared to the range of energies considered in this paper. Selection of polytype is therefore informed more by sample quality, mass, and ease of fabrication, as well as potential science reach, than by ultimate phonon resolution. The difference in science reach between the polytypes is the focus of the next sections of this paper.

Finally, we note that the quoted resolutions apply to readout limited by phonon dynamics, and implicitly assume that the phonon sensors used to read out these signals have a higher bandwidth than the crystal collection time. For these designs, a sensor with a response time of $\sim$1~\us~would be able to achieve within a factor of a few of these projected resolutions. This requirement, and the additional requirement that sensors be individually read out for design D, suggests that superconducting resonator technologies, such as Kinetic Inductance Detectors (KIDs), or switching technologies, such as superconducting nanowires, are more likely to be the technology of choice than TESs or thermal sensors. The former technologies have noise temperature and response time that are independent of thermal conductance, and are intrinsically multiplexable. The development of faster, low-$T_c$ TES detectors which are capable of frequency-domain multiplexing would allow them to be competitive at the lower thresholds quoted here.

\section{Theoretical Framework}\label{sec:theoreticalFramework}

We now move to describing the DM frameworks that can be detected via SiC detectors.  
We consider the following possible signals from sub-GeV DM interactions: scattering off nuclei elastically, scattering into electron excitations for $m_\chi \gtrsim \MeV$, phonon excitations for $\keV \lesssim m_\chi \lesssim 10\, \MeV$, and absorption of dark matter into electronic and phonon excitations for $10\, \meV \lesssim m_\chi \lesssim 100\, \eV$. In all cases, $\rho_\chi=0.3\;{\rm GeV}/{\rm cm}^3$ is the DM density, and $f_\chi({\bf v})$ is the DM velocity distribution, which we take to be the Standard Halo Model~\cite{PhysRevD.33.3495} with $v_0 = 220$ km/s, $v_{\rm Earth} = 240$ km/s, and $v_{\rm esc} = 500$ km/s.
\subsection{Elastic DM-nucleus scattering}

Assuming spin-independent interactions, the event rate  from dark matter scattering off of a nucleus in a detector of mass $m_{\rm det}$ is given by the standard expression~\cite{lewinsmith}
\begin{equation}
\frac{dR}{dE_r} = \frac{ m_{\rm det} \rho_{\chi}\sigma_0}{2m_{\chi}\mu_{\chi N}^2}F^2(q) F^2_{\rm med}(q) \int_{v_{\rm min}} \frac{f_\chi({\bf v})}{v}d^3 {\bf v}.
\end{equation}
Here $q = \sqrt{2 m_T E_r}$ is the momentum transfer, $m_T$ is the target mass, $m_\chi$ is the DM mass, $\mu_{\chi N}$ is the reduced mass of the DM-nucleus system, $E_r$ is the recoil energy, $F(E_r)$ is the nuclear form factor of DM-nucleus scattering (we adopt the Helm form factor as in Ref.~\cite{lewinsmith}), and the form factor  $F^2_{\rm med}(q)$ captures the form factor for mediator interactions ({\it  i.e}., long-range or short-range). The cross-section $\sigma_0$ is normalized to a target nucleus, but to compare different media, this cross-section is re-parameterized as~\cite{lewinsmith,hertel}
\begin{equation}
\sigma_0 = A^2\left(\frac{\mu_{\chi N}}{\mu_{\chi n}}\right)^2\sigma_{n},
\end{equation}
where $A$ is the number of nucleons in the nucleus, and $\mu_{\chi n}$ is the DM-nucleon reduced mass. 

For a sub-GeV dark matter particle, we have $\mu_{\chi N}\rightarrow m_{\chi}$, $\sigma_0\rightarrow A^2\sigma_n$, and $F(E_r)\rightarrow 1$, such that
\begin{equation}
\frac{dR}{dE_r} \approx m_{\rm det}\frac{\rho_{\chi}A^2\sigma_n}{2m_{\chi}^3} F^2_{\rm med}(q)  \int_{v_{\rm min}}\frac{f_\chi( {\bf v})}{v}d^3 {\bf v},
\end{equation}
which would seem to imply that a heavier nucleus is always more sensitive to dark matter from a pure event-rate perspective. Hidden in the integral, however, is the fact that
\begin{equation}
v_{\rm min} = \sqrt{\frac{E_r(m_{\chi}+m_T)}{2\mu_{\chi N}m_{\chi}}} \rightarrow \sqrt{\frac{E_r m_{T}}{2m_{\chi}^2}}
\end{equation}
in this limit, which implies scattering off of heavier targets is kinematically suppressed. 

For heterogeneous targets, the standard modification to this rate formula is to weight the event rate for a given atom by its fractional mass density. 
For a SiC crystal of mass $m_{\rm det}$ and equal number density of Si and C nuclei, we have the total rate 
\begin{equation}
    \left(\frac{dR}{dE_r}\right)_{\rm SiC} = \frac{1}{2m_{\rm SiC}}\left[m_{\rm Si}\left(\frac{dR}{dE_r}\right)_{\rm Si}+m_{\rm C}\left(\frac{dR}{dE_r}\right)_{\rm C}\right]
    \nonumber
\end{equation}
where the rates for Si and C are computed for the given detector mass. This is a reasonable assumption for interactions in which the scattered DM particle only probes a single nucleus.   For sufficiently low $E_r$ comparable to the typical phonon energy, the assumption is no longer valid.  This can be seen from the fact that the interaction of DM with single or multi-phonons is an expansion in $q^2/(m_T \omega)$~\cite{trickle2019,Campbell-Deem:2019hdx}, so that we transition to the nuclear recoil regime when $E_r \gg \omega_{\rm phonon}$. 
In this paper we consider elastic nuclear recoils down to 0.5~eV, well above the energy at the highest optical phonon, and consider DM as acting locally on a single nucleus from the standpoint of the initial interaction. For energy depositions between the highest optical phonon energy, $\sim 120$ meV, and 0.5 eV, we expect the signal rate to be dominated by multiphonon interactions.

To compute NR limits, the behavior at low DM mass is strongly dependent on the energy threshold, while the high-mass behavior depends on the upper limit for accurate energy reconstruction. Athermal phonon calorimeters can provide very low thresholds but are intrinsically limited in dynamic range. To account for this, we assume 3 orders of magnitude in dynamic range, similar to what has been seen in detectors with ${\cal O}(\rm eV)$ thresholds~\cite{kurinsky}. This means that the upper integration limit is set to $10^3\sigma_{t}$, where the threshold $\sigma_t$ is assumed to be 5 times the resolution. 

\subsection{DM-phonon scattering \label{sec:phonon} }

The formalism to compute single phonon excitations from DM scattering was detailed previously in Refs.~\cite{Knapen:2017ekk,Griffin:2018bjn,trickle2019}. The scattering rate per unit time and per unit target mass can be written generally as
\begin{align}
    R = \frac{1 }{\rho_T} \frac{\rho_\chi}{m_\chi} \int d^3 {\bf v} f_\chi({\bf v}) \, \Gamma({\bf v}), \label{eq:general_rate}
\end{align}
where $\rho_T$ is the total target density. $\Gamma({\bf v})$ is the scattering rate per dark matter particle with velocity ${\bf v}$, given by
\begin{align}
    \Gamma({\bf v}) \equiv \frac{\bar \sigma_{\chi} }{4 \pi \mu_{\chi n}^2} \int \frac{d^3 \bfq}{\Omega} \, F_{\rm med}^2(q) \, S_{\rm med}({\bf q}, \omega).
    \label{eq:rate}
\end{align}
$\mu_{\chi n}$ is the DM-nucleon reduced mass and $\bar \sigma_\chi$ is a fiducial cross section which we will define later for specific models. $\Omega$ is the primitive cell volume, and can also be written as $(\sum_d m_d)/\rho_T$ where $d$ sums over all atoms in the cell. As above, the form factor  $F^2_{\rm med}(q)$ captures the form factor for mediator interactions ({\it i.e.}, long-range or short-range). Finally, the structure factor $S_{\rm med}({\bf q}, \omega)$ encapsulates the phonon excitation rate for a given momentum transfer $\bfq$ and energy deposition $\omega$; note that it depends on the mediator through its couplings to the nuclei and electrons in a given target. 

As specific examples, we first consider a mediator that couples to nuclei proportional to atomic number $A$, in which case
\small
\begin{align}
    S_{\rm med}(\bfq, \omega) = \sum_{\nu,\bfk, \bfG} \frac{\delta(\omega - \omega_{\nu, \bfk})}{2 \omega_{\nu, \bfk}}   |F_{N,\nu}(\bfq,\bfk)|^2 \delta_{\bfk - \bfq, \bfG}
    \label{eq:structure_factor}
\end{align}
\normalsize
where $\nu$ labels phonon branch and $\bfk$ denotes crystal momentum within the first Brillouin zone. The $\bfG$ are reciprocal lattice vectors, and for sub-MeV dark matter the ${\bfG =0}$ piece of the sum dominates. The phonon form factor for this mediator is
\small
\begin{align}
    |F_{N,\nu}(\bfq, \bfk)|^2 = \left| \sum_d \frac{A_d \,  \bfq \cdot {\bf e}_{\nu, d, \bfk}^*  }{\sqrt{m_d}}  e^{-W_d(\bfq)} \, e^{i(\bfq - \bfk)\cdot {\bf r}_d^0} \right|^2, 
\end{align}
\normalsize
where $d$ labels atoms in the primitive cell and ${\bf r}_d^0$ are the equilibrium atom positions. We determine the phonon eigenvectors, ${\bf e}_{\nu, d, \bfk}$, and band structure $\omega_{\nu,\bfk}$ numerically from first-principles calculations described later in this section. Finally, $W_d(\bfq)$ is the Debye-Waller factor, which we can approximate as $W_d(\bfq) \approx 0$ since the rates for sub-MeV DM are dominated by low $q$. With this phonon form factor, sub-MeV dark matter dominantly couples to longitudinal acoustic phonons.

We next consider a mediator that couples to electric charge, such as a dark photon mediator $A'$. The structure factor has the same form as in Eq.~\eqref{eq:structure_factor}, but with $F_{N,\nu}$ replaced by the phonon form factor
\small
\begin{align}
    |F_{A',\nu}({\bf q},{\bfk})|^2 = \left| \sum_d \frac{\bfq \cdot {\bf Z}_d^* \cdot {\bf e}^*_{\nu,d,{\bfk}} }{\epsilon_\infty \sqrt{m_d}} e^{-W_d(\bfq)} \, e^{i(\bfq - \bfk)\cdot {\bf r}_d^0} \right|^2, \nonumber
\end{align}
\normalsize
where we have assumed diagonal high-frequency dielectric constant $\epsilon_\infty$ and where ${\bf Z}_d^*$ is the matrix-valued Born effective charge of atom $d$ in the unit cell. It is the nonzero Born effective charges in polar semiconductors that permits sensitivity to these models, and it has been found that the most important mode excitation is the highest energy longitudinal optical phonon mode.

\subsubsection{Daily modulation \label{sec:directional} }

The anisotropic crystal structures of SiC polymorphs imply a directional-dependence of DM-phonon scattering. As the Earth rotates, there is a corresponding modulation in the rate over a sidereal day, which can provide a unique discriminant for a DM signal in the event of a detection. This effect can be captured by accounting for the time-dependent direction of the Earth's velocity with respect to the lab frame in the DM velocity distribution, $f_\chi({\bf v})$.

This approach to calculating the directional signal was previously taken in Ref.~\cite{Griffin:2018bjn}, where it was computed for Al$_2$O$_3$ (sapphire) which has a rhombohedral lattice structure. The rate depends on the orientation of the crystal relative to the DM wind or equivalently Earth's velocity. Similar to Ref.~\cite{Griffin:2018bjn}, we choose the crystal orientation such that the $z$-axis is aligned with the Earth's velocity at $t=0$. Since the Earth's rotation axis is at an angle of $\theta_e \approx 42^\circ$ relative to the Earth's velocity, at time $t=1/2$ day, the $z$-axis of the crystal will be approximately perpendicular to the DM wind. For the rhombohedral and hexagonal lattice structures, the convention is that the $z$-axis corresponds to the primary crystal axis, and so we expect that this configuration should give a near-maximal modulation rate.

\subsection{DM-electron scattering}

Eqs. \eqref{eq:general_rate} and \eqref{eq:rate} are applicable to electron scattering as well, with the appropriate substitutions. The structure factor $S({\bf q},\omega)$ for electron recoil is given by \cite{trickle2019,Essig:2015cda}
\begin{align}
S(\bm{q},\omega) & = 2 \sum_{i_1,i_2} \int_{BZ} \frac{d^3k~d^3k'}{(2\pi)^6} 2\pi \delta(E_{i_2,\bm{k}'} - E_{i_1,\bm{k}} - \omega) \times \nonumber
 \\ 
&\sum_{\bm{G}} (2\pi)^3 \delta(\bm{k}' - \bm{k} + \bm{G} - \bm{q}) |f_{[i_1\bm{k},i_2\bm{k}',\bm{G}]}|^2 
\end{align}
where $E_{i,\bm{k}}$ is the energy of a electron in band $i$ with crystal momentum $\bm{k}$ and ${\bf G}$ are the reciprocal lattice vectors. The crystal form factor $f_{[i_1\bm{k},i_2\bm{k}',\bm{G}]}$ is given by
\begin{align}
f_{[i_1\bm{k},i_2\bm{k}',\bm{G}]} = \sum_{\bm{G}'} u_{i_1}^*(\bm{k'}+\bm{G}+\bm{G'})u_{i_2}(\bm{k}+\bm{G'})
\end{align}
where $u_i(\bm{k})$ are the electron wavefunctions written in plane wave basis and normalized such that
\begin{align}
    \sum_{\bm{G}} |u_i(\bm{k}+{\bm{G}})|^2 = 1\,.
\end{align}

In our calculation of the electron recoil limits, we make the isotropic approximation following the formalism outlined in Ref. \cite{Essig:2015cda}. The scattering rate per unit time and per unit target mass is then simplified to
\small
\begin{align}\label{eq:Rerecoil}
    R = \frac{\bar{\sigma}_e}{2 \rho_T \mu^2_{\chi e}} \frac{\rho_\chi}{m_\chi} \int q dq d\omega \, F_{\rm med}^2(q) S(q,\omega)\eta(v_{\rm min}(q,\omega)) 
\end{align}
\normalsize
where $\mu_{\chi e}$ is the reduced mass of the DM and electron, and the integrated dark matter distribution $\eta(v_{\rm min})$ is given as in Ref.~\cite{Essig:2015cda}. The reference cross section $\bar\sigma_e$ is at a fixed reference momenta, which will be taken as $\alpha m_e$, with $\alpha$ the fine structure constant and $m_e$ the electron mass.
Results for daily modulation and thus directional detection of electron recoil signals in SiC will be presented in future work~\cite{future}.


\subsection{Absorption of sub-keV DM \label{sec:absorption} }

For a number of models, the bosonic DM absorption rate can be determined in terms of the conductivity of the material and photon absorption rate. Then the absorption rate is given as
\begin{align}
	R = \frac{1}{\rho_T} \frac{\rho_\chi }{m_\chi}  g_{\rm eff}^2 \sigma_{\rm 1}(m_\chi)
	\label{eq:rate_absorb}
\end{align}
where $\sigma_1(m_\chi)$ is real part of the optical conductivity $\hat\sigma$ of the material, namely the absorption of photons with frequency $\omega = m_{\chi}$, and $g_{\rm eff}$ is an effective coupling constant appropriate per  model~\cite{Hochberg:2016sqx,Griffin:2018bjn}, as will be detailed below.

The conductivity of the material can be obtained from measurements or by calculation. For $m_\chi $ greater than the electron band gap, we use measurements on amorphous SiC thin films from Ref.~\cite{SiCdata}. This data does not capture the differences between polymorphs of SiC, with band gaps ranging from 2.36 eV to 3.25 eV for those considered here, but we expect the differences to be small for $m_\chi$ well above the electron band gap. 

For $m_\chi$ below the electron band gap, absorption can occur into single optical phonons as well as multi-phonons. In this case, there is limited data or calculation available for sub-Kelvin temperatures. To gain further insight, we can use an analytic approximation for the dielectric function~\cite{Griffin:2018bjn}:
\begin{align}
 	\hat \epsilon(\omega) = \epsilon_\infty \prod_\nu \frac{\omega_{{\rm LO},\nu}^2 - \omega^2 + i \omega \gamma_{{\rm LO},\nu}}{\omega^2_{{\rm TO},\nu} -\omega^2 +  i \omega \gamma_{{\rm TO},\nu}},
	\label{eq:permittivity}
\end{align}
with a product over all optical branches, and where $\gamma$ is the phonon linewidth, and TO (LO) abbreviate transverse (longitudinal) optical phonons. The dielectric function is related to the complex conductivity $\hat \sigma(\omega)$ by $\hat \epsilon(\omega) = 1 + i \hat \sigma/\omega$. 

We separately consider the conductivity parallel to the c-axis, $\hat \epsilon_\parallel(\omega)$, and perpendicular to the c-axis, $\hat \epsilon_\perp(\omega)$. In SiC, there is a strong optical phonon branch for each of these directions, corresponding to the highest energy optical phonons ($A_1$ in the parallel direction, $E_1$ in the perpendicular direction)~\cite{mutschke1999infrared}. For these phonons, the LO and TO frequencies are compiled in Ref.~\cite{mutschke1999infrared}, where the values are nearly identical across all polymorphs. Because there are very limited low-temperature measurements of the phonon linewidths, we use $\gamma_{\rm LO} = 2.6/\textrm{cm}$ and $ \gamma_{\rm TO} = 1.2/\textrm{cm}$ in all cases. These values come from our calculations of the linewidth of the optical phonons in the 3C polymorph, and are also in agreement with experimental data~\cite{3Clinewidths}. The calculation of  linewidths is discussed in Appendix \ref{app:first_principles_calcs}.

Here we only consider the absorption into the strongest phonon branch for the parallel and perpendicular directions. Accounting for the fact that the DM polarization is random, we will take an averaged absorption over these phonon modes, $\langle R \rangle = \frac{2}{3}R_\perp + \frac{1}{3}R_{\parallel}$. With the above approximations, we find that the absorption rate for the strongest mode is nearly identical across all polymorphs.
However, depending on the polymorph, there are additional lower energy optical phonons with weaker absorption, and which can have large mixing of transverse and longitudinal polarizations. Furthermore, there is absorption into multiphonons. While these contributions are not included in our analytical computation, we expect the qualitative behavior of the low-mass absorption rate to be well captured by the range formed by the available measurements from Ref.~\cite{SiCdata} and the above calculation.

\section{Results}\label{sec:results}

\subsection{DM with scalar nucleon interactions}

\begin{figure*}[t!]
\begin{center}
\includegraphics[width=0.98\textwidth]{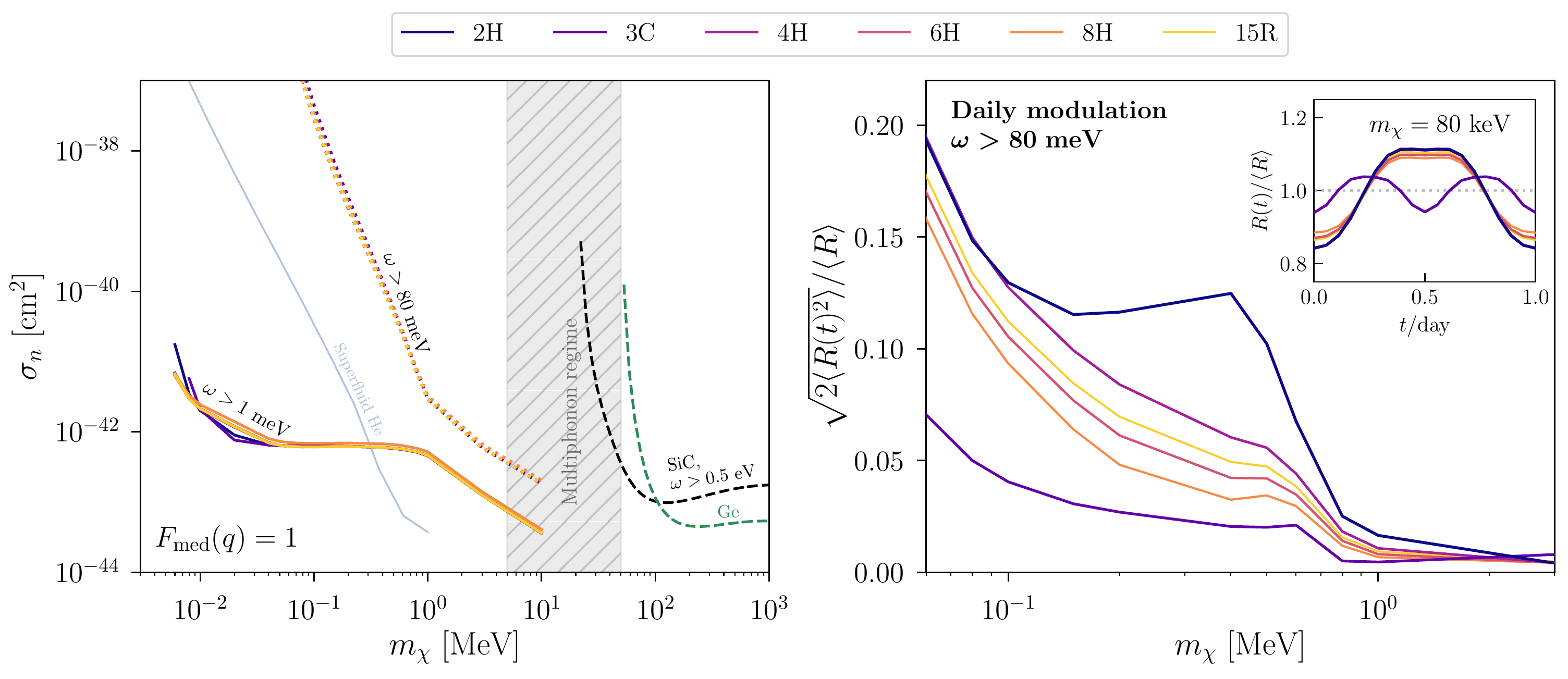}
\caption{ \label{fig:neutron_sigma_massive} Reach and daily modulation for DM interactions mediated by a scalar coupling to nucleons, assuming a massive mediator.  {\bf Left:} All reach curves are obtained assuming kg-year exposure and zero background. For single phonon excitations relevant for $m_\chi \lesssim 10$ MeV, we show two representative thresholds of 1~meV (solid lines) and 80~meV (dotted) for the different SiC polytypes. We also show the reach for a superfluid He target~\cite{Knapen:2016cue}. The dashed lines show sensitivity to nuclear recoils assuming threshold of 0.5 eV. In the shaded region, it is expected that the dominant DM scattering is via  multiphonons (see discussion in Refs.~\cite{Campbell-Deem:2019hdx,trickle2019}). {\bf Right:} The daily modulation of the DM-phonon scattering rate as a function of DM mass, where the quantity shown corresponds exactly to the modulation amplitude for a purely harmonic oscillation. The modulation is much smaller for scattering into acoustic phonons $\omega > 1$ meV, so we only show scattering into optical phonons with $\omega > 80$ meV. The modulation amplitude is generally largest for 2H and smallest for 3C.  The inset compares the phase of the modulation among the polymorphs for $m_\chi$ = 80 keV.
}
\end{center}
\end{figure*}

\begin{figure*}[t!]
\begin{center}
\includegraphics[width=0.98\textwidth]{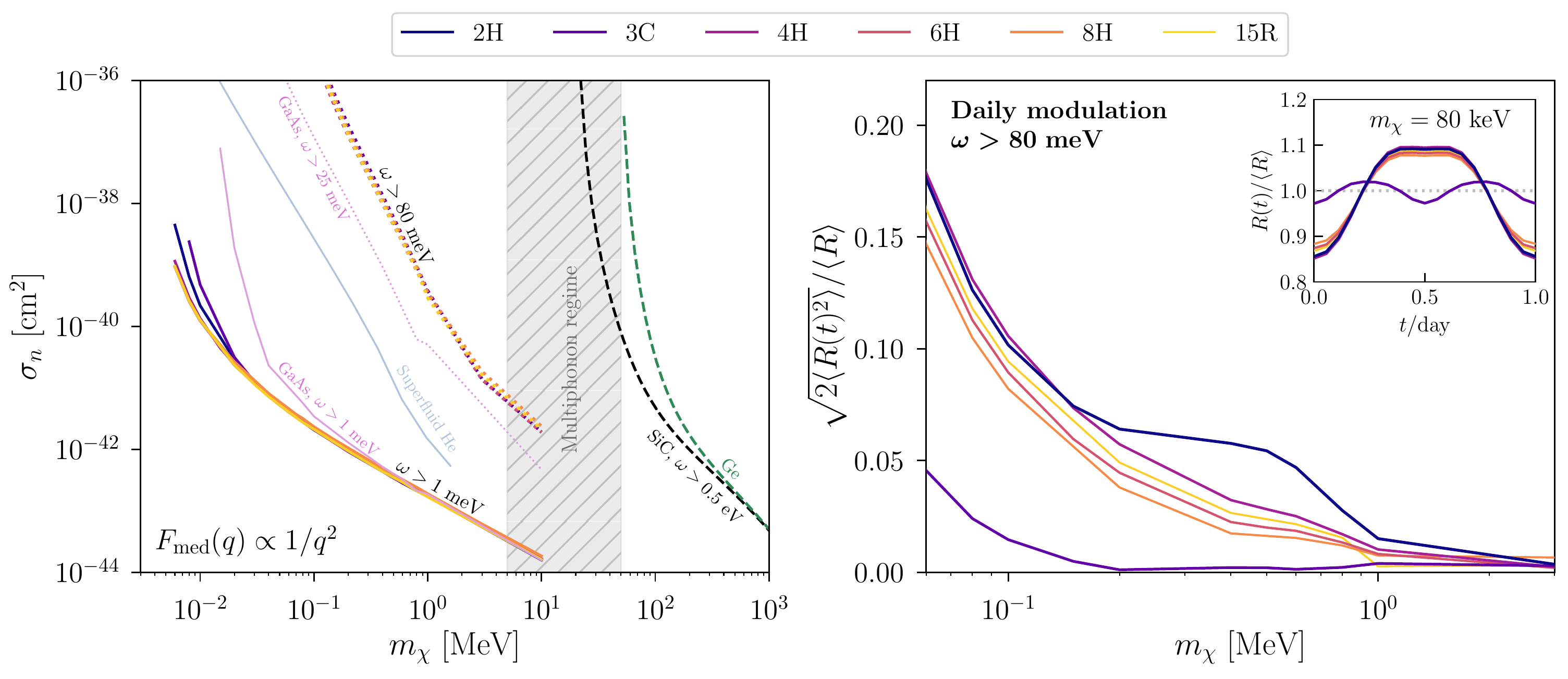}
\caption{ \label{fig:neutron_sigma_massless} Similar to Fig.~\ref{fig:neutron_sigma_massive}, but for DM interactions mediated by a massless scalar coupling to nucleons. In this case, we also compare with the reach of another polar material, GaAs, for acoustic and optical branch thresholds~\cite{Griffin:2018bjn}.
}
\end{center}
\end{figure*}

For dark matter with spin-independent scalar interactions to nucleons, we consider both the massive and massless mediator limit, corresponding to different choices of mediator form factor $F^2_{\rm med}(q)$. A discussion of the astrophysical and terrestrial constraints on both cases can be found in Ref.~\cite{Knapen:2017xzo}.

For the massive scalar mediator coupling to nucleons, the form factor is $F^2_{\rm med}(q) = 1$. The sensitivity of SiC to this model is shown in the left panel of Fig.~\ref{fig:neutron_sigma_massive} for the various SiC polytypes and also a few different experimental thresholds. For energy threshold $\omega > 0.5$ eV, we show the reach for nuclear recoils in a SiC target and compare with a representative target containing heavy nuclei.

The DM-phonon rate is determined using Eq.~\eqref{eq:rate}, where the fiducial cross section is $\bar \sigma_\chi \equiv \sigma_n$ and $\sigma_n$ is the DM-nucleon scattering cross section. With a threshold $\omega > $~meV, it is possible to access DM excitations into single acoustic phonons, which provide by far the best sensitivity. While this threshold would be challenging to achieve, we show it as a representative optimistic scenario where access to single acoustic phonons is possible.  The reach here is primarily determined by the speed of sound~\cite{Campbell-Deem:2019hdx}, and is thus fairly similar for all crystal structures. For comparison with additional polar crystal targets, see Ref.~\cite{Griffin:2019mvc}.

When the threshold is $\omega \gtrsim 20-30$ meV, the only excitations available are optical phonons. For DM which couples to mass number, there is a destructive interference in the rate to excite optical phonons, resulting in significantly worse reach~\cite{Knapen:2017ekk,Cox:2019cod}. In Fig.~\ref{fig:neutron_sigma_massive}, we also show a representative optical phonon threshold of  $\omega > 80$ meV  as this is just below the cluster of optical phonons of energy $90-110$ meV present in all polymorphs (see Fig.~\ref{fig:SiC_phonons}). Note that the reach for $\omega > 30$ meV is not significantly different from $\omega > 80$ meV, due to the destructive interference mentioned above.

While the optical phonon rate is much smaller than the acoustic phonon rate, the same destructive interference allows for a sizeable directionality in the DM scattering rate, and thus daily modulation. The right panel of Fig.~\ref{fig:neutron_sigma_massive} gives the daily modulation for DM scattering into optical phonons with threshold $\omega > 80$ meV. We find that the lowest modulation is for the 3C polytype, as expected given its higher degree of symmetry, and the largest modulation can be found in the 2H polytype. While the other polytypes of SiC can give comparable modulation to 2H, they contain many more phonon branches, which can wash out the signal. We also note that the modulation could be even larger with a lower threshold on the optical phonons, which was the case for sapphire in Ref.~\cite{Griffin:2018bjn}. However, if the threshold is reduced all the way to $\omega > $ meV such that acoustic phonons are accessible, the modulation is much smaller. 

In the massless mediator limit, we assume dark matter couples to nucleons through a scalar with mass $m_\phi \ll m_\chi v \sim 10^{-3} m_\chi$. For sub-MeV DM, constraints on this model are much less severe than in the heavy mediator case~\cite{Knapen:2017xzo}. Then we can approximate the DM-mediator form factor as
\begin{align}
    F^2_{\rm med}(q) = \left( \frac{q_0}{q} \right)^4
\end{align}
where $q_0 = m_{\chi} v_0$ is a reference momentum transfer.
In this case $\sigma_n$ is a reference cross section for DM-nucleon scattering with momentum transfer $q_0$. 
The projected sensitivity to the massless mediator model from single-phonon excitations in SiC is shown in the left panel of Fig.~\ref{fig:neutron_sigma_massless}. Here we also show the reach for a GaAs target, which has a lower sound speed and thus more limited reach at low DM mass~\cite{Griffin:2018bjn}. For comparison with additional polar crystal targets, see Ref.~\cite{Griffin:2019mvc}.

The daily modulation amplitude for a massless scalar mediator is shown in the right panel of Fig.~\ref{fig:neutron_sigma_massless}. Similar to the massive mediator case, we only have a sizeable modulation for scattering into optical phonon modes, and find that 2H (3C) tends to give the largest (smallest) amplitude.

We conclude with a brief discussion of how SiC compares with other commonly considered target materials for DM with scalar nucleon interactions. Because SiC has a high sound speed similar to that of diamond, the sensitivity to acoustic phonon excitations extends to lower DM mass than in Si, Ge, or GaAs. Furthermore, depending on the polytype of SiC, the daily modulation in SiC is expected to be much larger than Si, Ge, GaAs and diamond. The latter materials have cubic crystal structures where the atoms in a unit cell have identical or very similar mass, so we expect the modulation to be similar to that of GaAs, found to be sub-percent level in Ref.~\cite{Griffin:2018bjn}. In terms of both reach and directionality, SiC is perhaps most similar to sapphire, and has advantages over many other well-studied target materials.

\subsection{DM-electron interactions \label{sec:DMelectron_result} }

\begin{figure*}[t!]
\begin{center}
\includegraphics[width=0.49\textwidth]{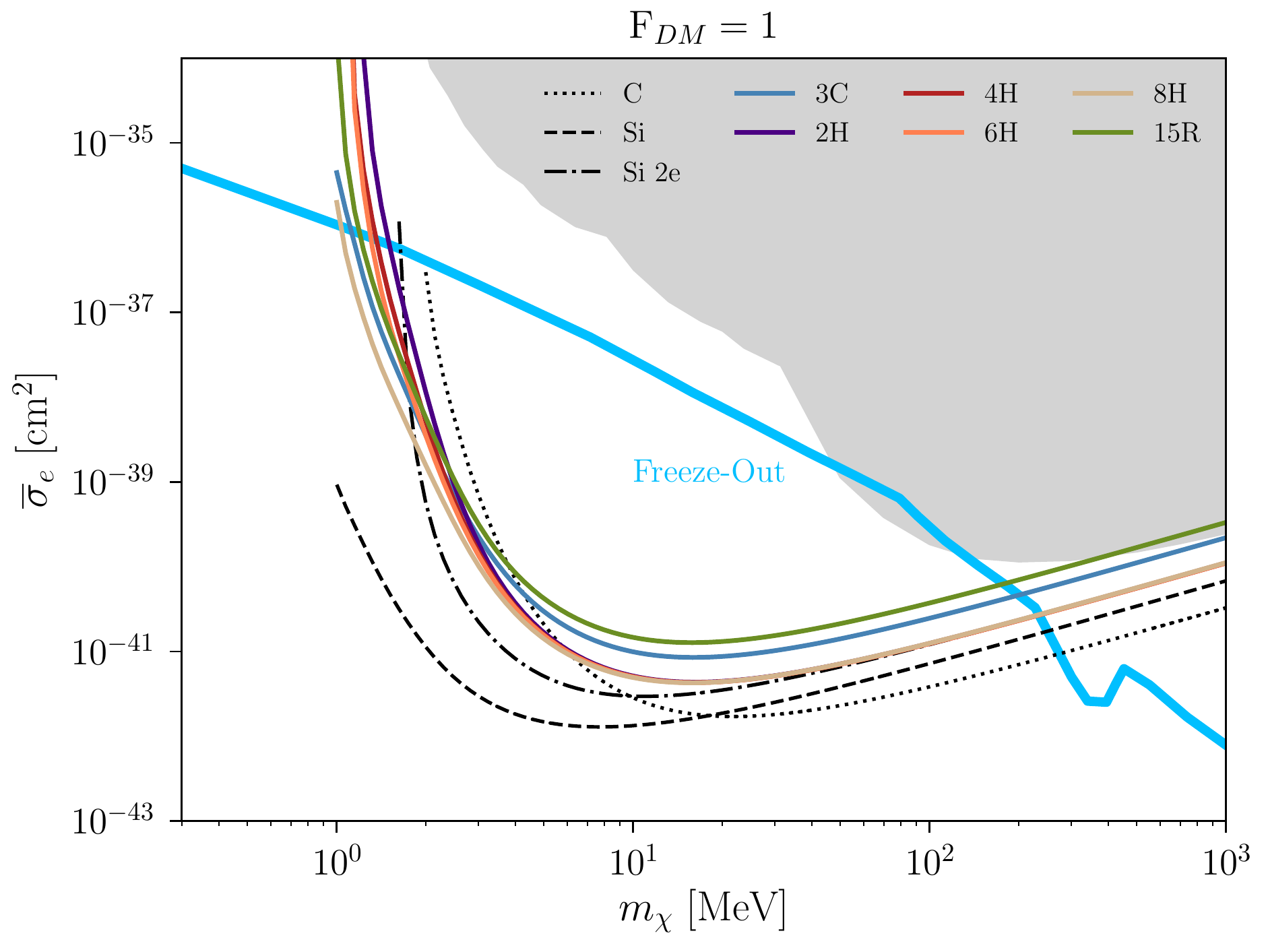}
\includegraphics[width=0.49\textwidth]{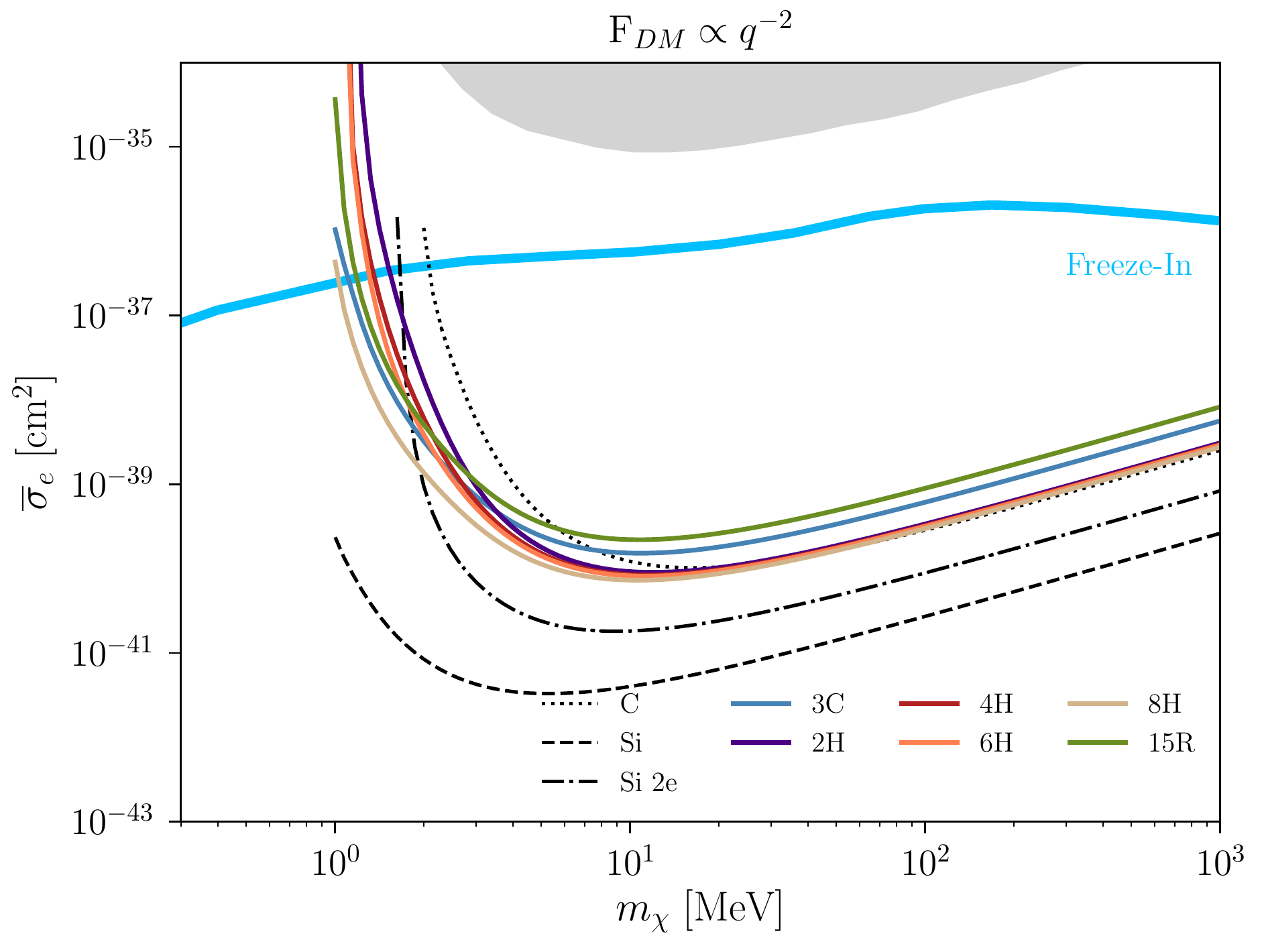}
\caption{ The reach into DM-electron scattering parameter space for 1 kg-year of exposure of select polytypes of SiC for  heavy~({\bf left}) and light~({\bf right}) mediators. For comparison, we also show the reach of Si and diamond, assuming a threshold of one electron or energy sensitivity down to the direct band gap in the given material. The reach of Si given a 2-electron threshold is shown for comparison, for the case that charge leakage substantially limits the reach at 1-electron. Relic density targets from Ref.~\cite{CosmicVisions} are shown as thick blue lines for the freeze-in and freeze-out scenarios respectively. The grey shaded region includes current limits from SENSEI~\cite{barak2020sensei}, SuperCDMS HVeV~\cite{HVeV2020}, DAMIC~\cite{DAMIC_ERDM}, Xenon10~\cite{Essig:2017kqs}, Darkside~\cite{DarksideER}, and Xenon1T~\cite{Aprile:2019xxb}. \label{fig:electronRecoil}}
\end{center}
\end{figure*}

We now present our results for DM that scatters with electrons through exchange of a scalar or vector mediator (that is not kinetically mixed with the photon).  Our results for the reference cross section $\bar \sigma_e$ of Eq.~\eqref{eq:Rerecoil} are given in Fig.~\ref{fig:electronRecoil} for the heavy ({\it left}) and light ({\it right}) mediator cases, with form factors
\begin{align}
    F_{\rm med}^2(q) = \begin{cases}
    1 & \textrm{ heavy mediator} \\
    (\alpha m_e)^4/q^4 & \textrm{ light mediator}
    \end{cases}
    \label{eq:F_med_ER}
\end{align}
For comparison, we also show the reach of Si and diamond. Thick blue curves indicate relic density targets from Ref.~\cite{CosmicVisions}. The grey shaded region show existing limits from SENSEI~\cite{barak2020sensei}, SuperCDMS HVeV~\cite{HVeV2020}, DAMIC~\cite{DAMIC_ERDM}, Xenon10~\cite{Essig:2017kqs}, Darkside~\cite{DarksideER} and Xenon1T~\cite{Xenon1T}. 

The results of Fig.~\ref{fig:electronRecoil} show that the reach of SiC to DM-electron scattering is similar to that of diamond  at high mass for the case of a light mediators, and comparable to the silicon two-electron reach for the heavy mediator case. The relation of the reach between SiC polytypes is similar to that found in Figs.~\ref{fig:neutron_sigma_massive} and~\ref{fig:neutron_sigma_massless}, in that the majority of the difference at low-mass can be attributed to the different band gaps. We do observe, however, that the reach of 3C at high mass is roughly half an order of magnitude less than the hexagonal polytypes, despite having the smallest band gap. This can be understood by noticing that the density of state near the conduction band minima is smaller than that in the other polytypes, thus limiting the available phase space. The reach of 15R is also significantly worse than the other polytypes because the size of its small Brillouin zone is poorly matched to the typical momentum transfer (few keV).

We learn that SiC can probe DM-electron scattering processes in a complementary manner to silicon and diamond. As mentioned earlier, prospects for directional detection of electron recoil signals in the various polytypes of SiC will be described in future work~\cite{future}.

\subsection{DM with dark photon interactions}

\begin{figure*}[t!]
\begin{center}
\includegraphics[width=0.98\textwidth]{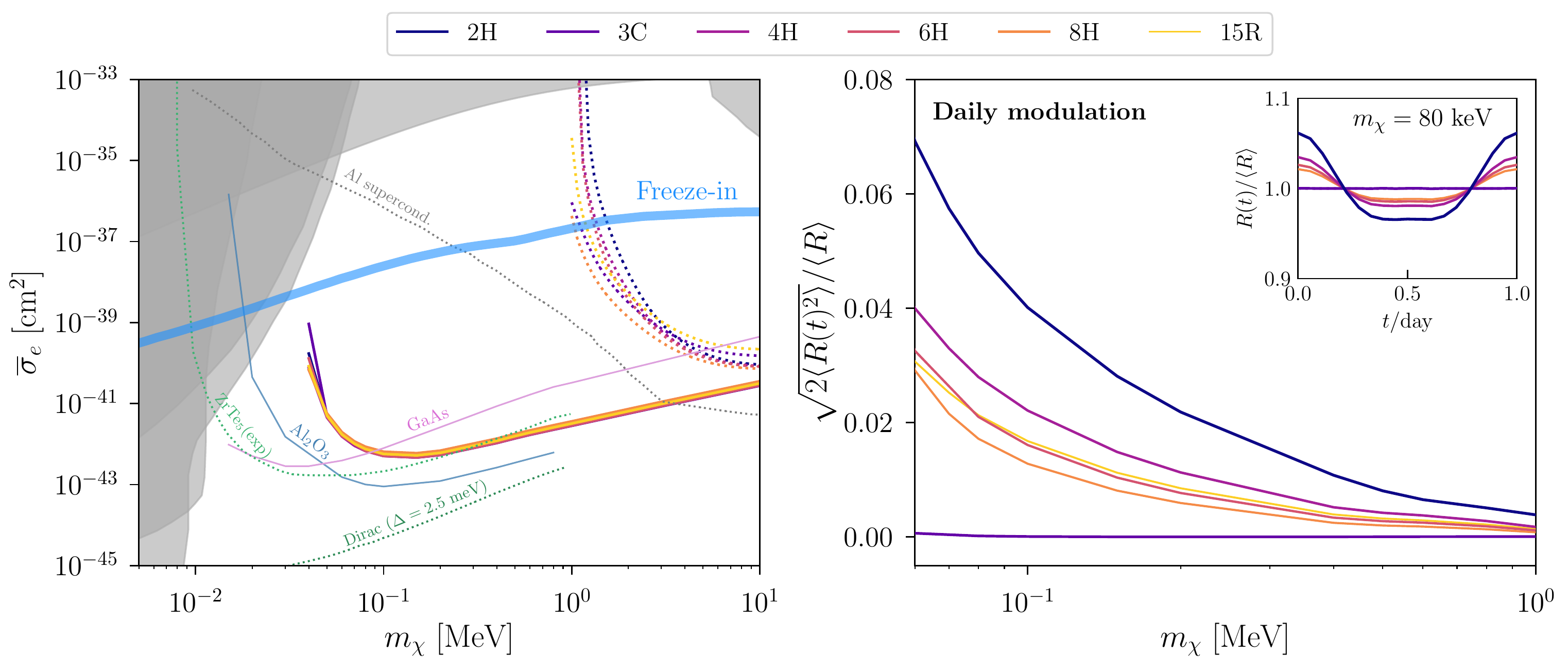}
\caption{ \label{fig:darkphoton} Reach and daily modulation for DM-phonon interactions mediated by a massless dark photon. {\bf Left:} The reach is shown assuming kg-year exposure and zero background. The reach from single optical phonon excitations in SiC (solid lines) is similar for all the polytypes, while the dotted lines in same colors are the electron recoil reach from Fig.~\ref{fig:electronRecoil}. The thick solid blue line is the predicted cross sections if all of the DM produced by freeze-in interactions~\cite{Essig:2011nj,Dvorkin:2019zdi}, and the shaded regions are constraints from stellar emission~\cite{Vogel:2013raa,Chang:2018rso} and Xenon10~\cite{Essig:2017kqs}.  We also show the reach from phonon excitations in other polar materials, GaAs and Al$_2$O$_3$~\cite{Knapen:2017ekk,Griffin:2018bjn}, and from electron excitations in an aluminum superconductor~\cite{Hochberg:2015fth} and in Dirac materials, shown here for the examples of ZrTe$_5$ and a material with gap of $\Delta = 2.5$ meV~\cite{Hochberg:2017wce}. (For clarity, across all materials, all electron recoil curves are dotted and all phonon excitation curves are solid.) {\bf Right:} The daily modulation of the DM-phonon scattering rate as a function of DM mass, where the quantity shown corresponds exactly to the modulation amplitude for a purely harmonic oscillation. The modulation is negligible in the 3C polytype due to its high symmetry, and is largest in 2H. The inset compares the phase of the modulation among the polytypes for $m_\chi$ = 80 keV. }
\end{center}
\end{figure*}

We now consider a DM candidate with mass $m_\chi$ that couples to a dark photon $A'$ of mass $m_{A'}$, where the dark photon has a kinetic mixing $\kappa$ with the Standard Model photon,
\begin{equation}
    {\cal L}\supset -\frac{\kappa}{2} F_{\mu\nu} F'^{\mu\nu}\,.
\end{equation} 
We again take two representative limits of this model: scattering via a massive or nearly massless dark photon. 

For a massive dark photon, the electron-scattering cross section $\bar \sigma_e$ in terms of model parameters is 
\begin{align}
    \bar \sigma_e = \frac{16 \pi \, \kappa^2 \alpha_\chi \alpha \, \mu_{\chi e}^2 }{\left[(\alpha m_e)^2 + (m_{A'})^2\right]^2}
\end{align}
and the DM-mediator form factor is $F_{\rm med}^2(q) = 1$. For the parameter space below $m_\chi \approx $ MeV, there are strong astrophysical and cosmological constraints~\cite{Essig:2015cda,Knapen:2017xzo} and the reach from exciting optical phonons is limited, so we do not consider DM-phonon scattering.  The electron scattering reach is the same as the heavy mediator limit of the previous section, shown in the left panel of Fig.~\ref{fig:electronRecoil}.

For a nearly-massless dark photon, we consider both electron recoils and optical phonon excitations.  Optical phonons are excited through the mediator coupling to the ion (nucleus and core electrons), which is given in terms of the Born effective charges discussed in Section~\ref{sec:phonon}. For comparison with the literature, we will show both the electron recoil and optical phonon reach in terms of the electron-scattering cross section. 
This electron-scattering cross section is defined at a reference momentum transfer, given in terms of the dark fine structure constant $\alpha_\chi = g_\chi^2/(4\pi)$:
\begin{align}
    \bar \sigma_e = \frac{16 \pi \, \kappa^2 \alpha_\chi \alpha \, \mu_{\chi e}^2 }{(\alpha m_e)^4}
\end{align}
where $\alpha$ is the fine structure constant and $\mu_{\chi e}$ is DM-electron reduced mass. As a result, for phonon scattering, the relevant cross section $\bar \sigma_\chi$ in Eq.~\eqref{eq:rate} is
\begin{align}
    \bar \sigma_\chi \equiv \frac{\mu_{\chi n}^2}{\mu_{\chi e}^2}  \bar \sigma_e.
\end{align}
The DM-mediator form factor for both electron and phonon scattering is
\begin{align}
    F_{\rm med}^2(q) = \left( \frac{\alpha m_e}{q} \right)^4.
\end{align}

The reach for different polytypes of SiC to the light mediator limit of this model is shown in the left panel of Fig.~\ref{fig:darkphoton}. The reach for $m_\chi > $ MeV is from DM-electron scattering, and is the same as the light mediator limit of the previous section (shown in the right panel of Fig.~\ref{fig:electronRecoil}). Although there is an additional in-medium screening for dark photon mediators compared to Section~\ref{sec:DMelectron_result}, we expect this to be a small effect for a relatively high-gap material such as SiC.  The sensitivity of SiC for $m_\chi < $ MeV is from exciting optical phonons, and is very similar across all polytypes. This is because the DM dominantly excites the highest energy optical phonon~\cite{Griffin:2018bjn}, which has the largest dipole moment and has similar energy in all cases. Furthermore, the coupling of the DM to this phonon is characterized by an effective Fr\"{o}hlich coupling that depends only on the phonon energy, $\epsilon_\infty$, and $\epsilon_0$~\cite{Knapen:2017ekk}. Again, it can be seen in Table~\ref{tab:properties} that all of these quantities are quite similar across the different polytypes. For completeness, we also show existing constraints from stellar emission~\cite{Vogel:2013raa,Chang:2018rso} and Xenon10~\cite{Essig:2017kqs}; projections for other materials such as Dirac materials~\cite{Hochberg:2017wce}, superconductors~\cite{Hochberg:2015fth}, and polar materials~\cite{Knapen:2017ekk,Griffin:2018bjn}; and target relic DM candidate curves~\cite{Essig:2011nj,Dvorkin:2019zdi}.

There are larger differences between polytypes in directional detection, which depends on the details of the crystal structure. The results for DM-phonon scattering are provided in the right panel of Fig.~\ref{fig:darkphoton}. Similar to the case of DM with scalar nucleon interactions, we find that 3C has the smallest modulation due to its higher symmetry, while 2H has the largest modulation. 

Comparing with other proposed polar material targets such as GaAs and sapphire, the reach of SiC for dark photon mediated scattering does not extend as low in DM mass because of the higher LO phonon energy. However, the directional signal is similar in size to that of sapphire and substantially larger than in GaAs. For additional proposed experiments or materials that can probe this parameter space, see for example Refs.~\cite{Griffin:2019mvc,Berlin:2019uco,Coskuner:2019odd,Geilhufe:2019ndy}.

\subsection{Absorption of dark photon dark matter}

\begin{figure*}[t!]
\begin{center}
\includegraphics[width=0.48\textwidth]{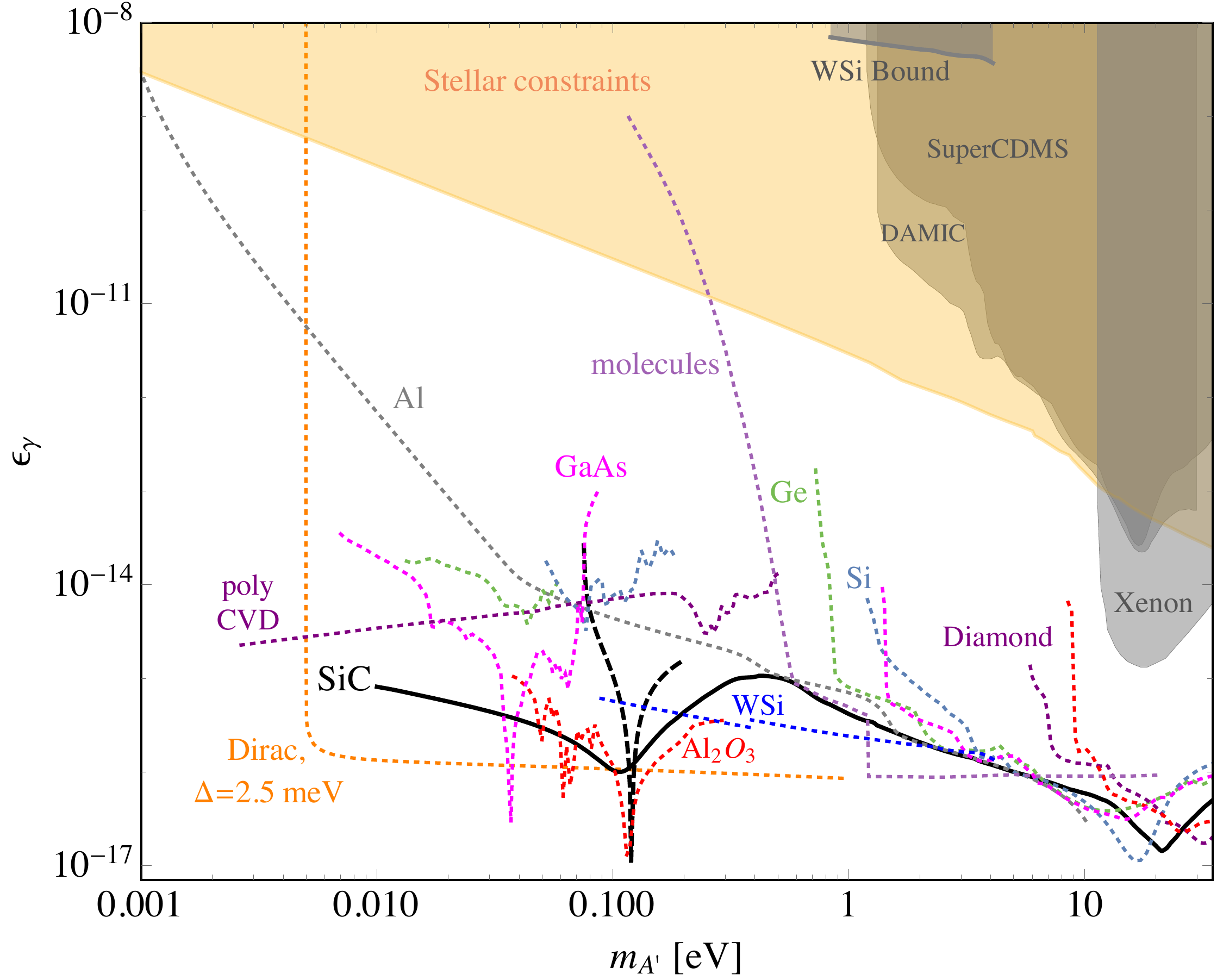}
\includegraphics[width=0.48\textwidth]{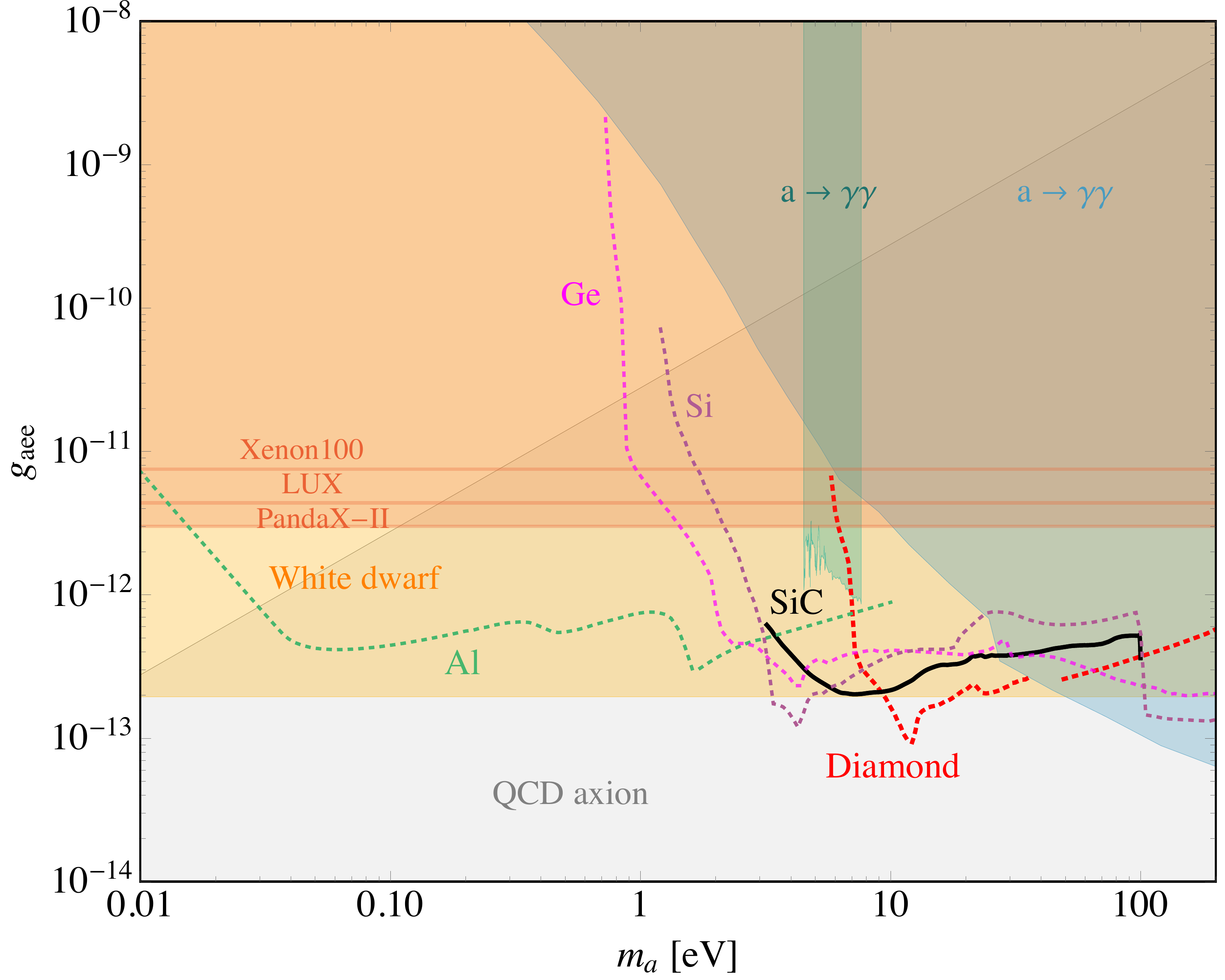}
\caption{ \label{fig:abs} 
 Absorption of kinetically mixed dark photons ({\it left}) and axion-like particles ({\it right}). {\bf Left:} Projected reach at 95\% C.L. for absorption of kinetically mixed dark photons. The expected reach for a kg-year exposure of SiC is shown by the solid and dashed black curves (using the data of Ref.~\cite{SiCdata} and strongest phonon branch, respectively). Projected reach for germanium and silicon~\cite{Hochberg:2016sqx}, diamond~\cite{diamonddetectors}, Dirac materials~\cite{Hochberg:2017wce}, polar crystals~\cite{Griffin:2018bjn}, molecules~\cite{Arvanitaki:2017nhi} superconducting aluminum~\cite{Hochberg:2016ajh} and WSi nanowire~\cite{Hochberg:2019cyy} targets are indicated by the dotted curves. Constraints from stellar emission~\cite{An:2013yua,An:2014twa}, DAMIC~\cite{Aguilar-Arevalo:2016zop}, SuperCDMS~\cite{Agnese:2018col} Xenon~\cite{An:2014twa} data and a WSi nanowire~\cite{Hochberg:2019cyy} are shown by the shaded orange, green, purple, light blue and blue regions, respectively. {\bf Right:} 
Projected reach at 95\% C.L. for absorption of axion-like particles. The reach of a kg-year exposure of SiC is shown by the solid black curve, where only excitations above the band gap are assumed.  The reach for semiconductors such as germanium and silicon~\cite{Hochberg:2016sqx}, diamond~\cite{diamonddetectors} and superconducting alumnium~\cite{Hochberg:2016ajh} targets is depicted by the dotted curves. Stellar constraints from Xenon100~\cite{Aprile:2014eoa}, LUX~\cite{Akerib:2017uem} and PandaX-II~\cite{Fu:2017lfc} data and white dwarfs~\cite{Raffelt:2006cw} are shown by the shaded red and orange regions. Constraints arising from (model-dependent) loop-induced couplings to photons are indicated by the shaded blue regions~\cite{Grin:2006aw,Arias:2012az}, while the QCD axion region is given in shaded gray.  }
\end{center}
\end{figure*}

Taking a dark photon with mass $m_{A'}$ and kinetic mixing $\kappa$ to be the dark matter candidate, the effective coupling $g_{\rm eff}^2$ in the absorption rate in Eq.~\eqref{eq:rate_absorb} must account for the in-medium kinetic mixing.  Thus we have $g_{\rm eff}^2 = \kappa_{\rm eff}^2$, with in-medium mixing of
\begin{align}
	\kappa_{\rm eff}^2 = \frac{\kappa^2 m_{A'}^4}{\left[m_{A'}^2 - \mbox{Re}~\Pi(\omega) \right]^2 + \mbox{Im}~\Pi(\omega)^2} = \frac{\kappa^2}{|\hat \epsilon(\omega) |^2}.
\end{align}
where $\Pi(\omega) = \omega^2(1 - \hat \epsilon(\omega))$ is the in-medium polarization tensor in the relevant limit of $|{\bf q}| \ll \omega$. 

The projected reach for absorption of SiC into the parameter space of kinetically mixed dark photons is shown in the left panel of Fig.~\ref{fig:abs}. As discussed in Section~\ref{sec:absorption},  we consider absorption into electron excitations using measurements of the optical conductivity of SiC from Ref.~\cite{SiCdata} (solid curve) as well as absorption into the strongest optical phonon mode for low masses (dashed curve). These black curves indicate the 95\% C.L. expected reach in SiC for a kg-year exposure, corresponding to 3 events. For comparison, we also show in dotted curves the projected reach of superconducting aluminum targets~\cite{Hochberg:2016ajh} and WSi nanowires~\cite{Hochberg:2019cyy}, semiconductors such as silicon, germanium~\cite{Hochberg:2016sqx} and diamond~\cite{diamonddetectors}, Dirac materials~\cite{Hochberg:2017wce}, polar crystals~\cite{Griffin:2018bjn} and molecules~\cite{Arvanitaki:2017nhi}. Stellar emission constraints~\cite{An:2013yua,An:2014twa} are shown in shaded orange, while the terrestrial bounds from DAMIC~\cite{Aguilar-Arevalo:2016zop}, SuperCDMS~\cite{Agnese:2018col}, Xenon data~\cite{An:2014twa} and a WSi superconducting nanowire~\cite{Hochberg:2019cyy} are shown in shaded gray. 
As is evident, SiC is a realistic target material that has prospects to probe deep into uncharted dark photon parameter space over a broad range of masses, from ${\cal O}(10\; {\rm meV})$ to 10's of eV.

\subsection{Absorption of axion-like particles}

Next we consider an axion-like particle (ALP) $a$ with mass $m_a$ that couples to electrons via 
\begin{equation}
	{\cal L}\supset \frac{g_{aee}}{2 m_e} (\partial_\mu a)\bar e \gamma^\mu \gamma^5 e\,.
\end{equation}
The absorption rate on electrons can be related to the absorption of photons via the axioelectric effect, and the effective coupling in Eq.~\eqref{eq:rateAxion} is then given by
\begin{equation}
\label{eq:rateAxion}
	g_{\rm eff}^2=  \frac{3 m_a^2}{4 m_e^2}  \frac{g_{aee}^2}{e^2}\,.
\end{equation}

Because the ALP directly couples to electrons, we consider only the absorption above the electron band gap. (Relating the couplings of the sub-gap phonon excitations is less straightforward due to the spin-dependence of the ALP coupling.) The projected reach for a kg-year exposure shown in the right panel of Fig.~\ref{fig:abs} by the solid black curve.  For comparison, we show the reach of superconducting aluminum~\cite{Hochberg:2016ajh} targets as well as silicon~\cite{Hochberg:2016sqx}, germanium~\cite{Hochberg:2016sqx} and diamond~\cite{diamonddetectors} by the dotted curves. Constraints from white dwarfs~\cite{Raffelt:2006cw}, Xenon100~\cite{Aprile:2014eoa}, LUX~\cite{Akerib:2017uem} and PandaX-II~\cite{Fu:2017lfc} are also shown. Constraints from the model-dependent loop-induced couplings to photons are indicted by shaded blue~\cite{Grin:2006aw,Arias:2012az}. The QCD axion region of interest is shown in shaded gray. We learn that SiC detectors can reach unexplored ALP parameter space complementary to stellar emission constraints.

\section{Discussion}

In this paper we proposed the use of SiC for direct detection of light DM. With  advantages over silicon and diamond--- including its polar nature and its many stable polymorphs---we have shown that SiC would serve as an excellent detector across many different DM channels and many mass scales: DM-nuclear scattering (direct and via single or multiple phonon excitations) down to ${\cal O}(10\, \rm keV)$ masses, DM-electron scattering down to ${\cal O}(10\, \rm MeV)$ masses, dark photon absorption down to ${\cal O}(10\, \rm meV)$ masses and axion-like absorption down to ${\cal O}(10\, \rm meV)$ masses, with prospects for directional detection as well. 

In particular, the high optical phonon energy in SiC (higher than that of sapphire) coupled with the high sound speed of all polytypes and long intrinsic phonon lifetime, makes SiC an ideal substrate for calorimetric phonon readout. There is substantial reach for dark photon coupled DM at higher energy thresholds than competing materials, and the presence of a strong bulk plasmon in SiC makes it a promising follow-up material for potential inelastic interactions of DM at the energy scales in the multi-phonon regime, as described in Refs.~\cite{kurinsky2020dark,Kozaczuk_2020}. 

In fact, since SiC exists in many stable polytypes, it allows us to compare the influence of crystal structure and hence bonding connectivity on their suitability as targets for various dark matter channels. Broadly, we see similar sensitivities and reach across the calculated polytypes as expected for a set of materials comprised of the same stoichiometric combination of elements. For DM-nucleon and DM-phonon interactions, we find very similar reach given the similar phonon spectra of the SiC polytypes. One difference is that polytypes with smaller unit cells will have the advantage of higher intrinsic phonon lifetimes, as the higher unit cell complexity will increase scattering.
More variation in reach among the polytypes, however, is found for DM-electron scattering due to the variation in electronic bandgaps across the SiC family. This trend in bandgap variation in SiC polytypes is well-discussed in the literature and is a result of the third nearest neighbor effects~\cite{Park1994}. We indeed see that, with increasing unit cell size, the decrease in bandgap in the H polytypes correspondingly leads to better reach, as expected. 

Materials-by-design routes explored for dark matter detection have focused on bandgap tuning~\cite{Inzani_et_al:2020}, and materials metrics for improved electron and phonon interactions~\cite{geilhufe2018materials,Griffin:2019mvc,geilhufe2020dirac,catena2020atomic}. A key advantage of SiC over other target proposals is its prospect for {\it directionality-by-design}---given the similar performance in reach across the polytypes, we can select a material that is optimized for directional detection. Our results indicate that, as expected, the highly symmetric cubic phase, 3C, exhibits no daily modulation, whereas the maximal modulation is achieved for the 2H phase. The 2H phase has inequivalent in-plane and out-plane crystallographic axes and so naturally has an anisotropic directional response. We further find that this effect is diminished for increasing the number of out-of-plane hexagonal units (decreasing the regularity of the unit cell) as the directional response becomes integrated out over repeated unit cells.    

As discussed earlier, one of the primary benefits of using SiC over other carbon-based crystals is the availability of large samples of the 4H and 6H polytypes; the 3C polytype is not currently at the same level of fabrication scale, and the 2H, 8H and 15R polytypes are scarce in the literature and not made in significant quantities. The charge mobility measurements for existing SiC samples indicate that purity of these crystals is not at the same level as comparable diamond and silicon, and there are few measurements of intrinsic phonon properties at cryogenic temperatures. In order to further develop SiC devices, studies of charge transport and phonon lifetime in a range of samples need to be undertaken so that current limitations can be understood and vendors can work to improve crystal purity. Device fabrication, on the other hand, is expected to be fairly straightforward due to past experience with basic metallization and the similarity of SiC to both diamond and Si. The availability of large boules of SiC, unlike for diamond, means that scaling to large masses for large detectors is much more commercially viable and cost effective.

The material response of the SiC polytypes should also be better characterized. In particular, studies of the non-ionizing energy loss of nuclear recoils needs to be modeled and characterized; photo-absorption cross-sections at cryogenic temperatures are needed, both above and below-gap; and the quantum yield of ionizing energy deposits needs to be better understood. SiC has already been shown to be much more radiation hard than Si, but more studies of radiation-induced defects will benefit both the use of SiC as a detector as well as a better understanding of vacancies used in quantum information storage. More practical studies of breakdown voltage and electron/hole saturation velocity will also inform detector modeling and readout.

\begin{acknowledgements}
We would like to thank Simon Knapen for early collaboration, and Rouven Essig and Tien-Tien Yu for clarifications regarding \texttt{QEDark}. We would also like to thank Lauren Hsu for feedback on an early paper draft. The work of YH is supported by the Israel Science Foundation (grant No. 1112/17), by the Binational Science Foundation (grant No. 2016155), by the I-CORE Program of the Planning Budgeting Committee (grant No. 1937/12), by the German Israel Foundation (grant No. I-2487-303.7/2017), and  by the Azrieli Foundation. TL is supported by an Alfred P. Sloan foundation fellowship and the Department of Energy under grant DE-SC0019195. Parts of this document were prepared by NK using the resources of the Fermi National Accelerator Laboratory (Fermilab), a U.S. Department of Energy, Office of Science, HEP User Facility. Fermilab is managed by Fermi Research Alliance, LLC (FRA), acting under Contract No. DE-AC02-07CH11359. TCY is supported by the U.S. Department of Energy under contract number DE-AC02-76SF00515. SMG and KI were supported by the Laboratory Directed Research and Development Program of LBNL under the DoE Contract No. DE-AC02-05CH11231. Computational resources were provided by the National Energy Research Scientific Computing Center and the Molecular Foundry, DoE Office of Science User Facilities supported by the Office of Science of the U.S. Department of Energy under Contract No. DE-AC02-05CH11231. The work performed at the Molecular Foundry was supported by the Office of Science, Office of Basic Energy Sciences, of the U.S. Department of Energy under the same contract.
\end{acknowledgements}

\appendix

\section{First-Principles Calculation Details of Electronic and Phononic Properties}
\label{app:first_principles_calcs}

Full geometry optimizations were performed using Density Functional Theory (DFT) with the Vienna \textit{Ab initio} Simulation Package (\textsc{vasp})~\cite{Kresse1993,Kresse1994,Kresse1996,Kresse1996a}, using projector augmented wave (PAW) pseudopotentials~\cite{Blo,Kresse1999} and the Perdew-Becke-Ernzerhof exchange-correlation functional revised for solids (PBEsol)~\cite{Perdew2008}. This gave lattice constants within 0.5\% of experimental values. The high frequency dielectric constants and Born effective charges were calculated using the density functional perturbation routines implemented in \textsc{vasp}. Force constants for generating the phonon dispersion spectra were calculated with the finite displacement method, using \textsc{vasp} and \textsc{phonopy}~\cite{Togo2015}.

The pseudopotentials used in the DFT calculations included \textit{s} and \textit{p} electrons as valence. A plane wave cutoff-energy of 800 eV was used with a $\Gamma$-centered \textit{k}-point grid with \textit{k}-point spacing $<$0.28 \AA$^{-1}$. This is equal to a $9\times9\times9$ grid for the 3C unit cell, and the equivalent \textit{k}-point spacing was used for the other polytypes and supercells. The cutoff-energy and \textit{k}-point spacing were chosen to ensure  convergence of the total energy to within 1 meV per formula unit and dielectric functions, Born effective charges and elastic moduli to within 1\%. The self-consistent field energy and force convergence criteria were $\SI{1E-8}{\electronvolt}$ and $\SI{1E-5}{\electronvolt\per\angstrom}$ respectively. For calculation of force constants within the finite displacement method, the following supercell sizes were used: $5\times5\times5$ for 3C (250 atoms), $4\times4\times4$ for 2H (256 atoms),
$3\times3\times3$ for 4H (216 atoms),
$3\times3\times3$ for 6H (324 atoms),
$3\times3\times3$ for 8H (432 atoms) and $3\times3\times3$ for 15R (270 atoms).

For phonon lifetimes and lidewidths, the \textsc{phono3py} code~\cite{phono3py} was used for calculation of phonon-phonon interactions within the supercell approach. Third-order force constants were were obtained from $4\times4\times4$ supercells for 3C and $3\times3\times3$ supercells for 2H. Lifetimes were computed on grids up to $90\times90\times90$. The acoustic lifetimes were averaged close to $\Gamma$ in the 0-2 THz frequency range. At 2 K the acoustic lifetimes are converged to an order of magnitude with respect to sampling grid. The optical lifetimes were averaged over all frequencies, and these converged to 3 significant figures. The corresponding optical linewidths were sampled close to $\Gamma$ for use in Eq.~\eqref{eq:permittivity}.

The electronic structures and wavefunction coefficients were also calculated by DFT, however the Heyd-Scuseria-Ernzerhof (HSE06) screened hybrid functional~\cite{Heyd2003a,Heyd2006} was used on PBEsol lattice parameters which gave excellent agreement with experimental band gaps (Table~\ref{tab:properties}). Band structures and isosurfaces were calculated with \textsc{vasp} using an increased \textit{k}-point density to ensure convergence of the band gap ($12\times12\times12$ grid for the 3C unit cell and equivalent for the other polytypes). For calculation of  the electron wavefunction coefficients, the Quantum Espresso code~\cite{Giannozzi2009,Giannozzi2017} was used to enable the use of norm-conserving Vanderbilt-type pseudopotentials~\cite{Hamann2013}. All other calculation choices between \textsc{vasp} and Quantum Espresso were kept consistent. The calculated detection rate was checked with respect to the \textit{k}-point density and plane wave energy cutoff of the underlying DFT calculations, and was found to be converged within 10\%. The following k-grids were used: $8\times8\times4$ for 2H, $8\times8\times8$ for 3C, $8\times8\times2$ for 4H, $8\times8\times2$ for 6H, $4\times4\times1$ for 8H, and $4\times4\times4$ for 15R. Bands up to 50 eV above and below the Fermi energy were used to evaluate the electron-DM matrix elements.

\section{Thermal Conductance and Phonon Lifetime}\label{app:kappa}

Here we show how thermal conductance measurements inform our estimations of phonon lifetime, which in turn are used in our discussion of phonon collection efficiency in Section~\ref{sec:phonon}. This model was used previously in Ref.~\cite{diamonddetectors} to estimate phonon lifetimes in diamond, but was not given explicitly in that paper.

Ref.~\cite{Callaway} defines the phonon relaxation time in the limit that $T\rightarrow 0$ as\footnote{The assumption of zero temperature allows us to ignore Umklapp processes, significantly simplifying this analysis.}
\begin{align}
    \tau_{r}^{-1} &= A\omega^4 +\frac{c_s}{L}+B_1T^3\omega^2 \\
    &= \frac{c_s}{L}\left[1+\frac{AL}{c_s}\omega^4+\frac{L}{c_s}B_1T^3\omega^2\right] \\
    &= \tau^{-1}_{b}\left[1+\left(\omega/\omega_{p}\right)^4+(\omega/\omega_B)^2\right]
\end{align}
where $L$ is the characteristic length scale of the crystal\footnote{For rough boundaries, L is simply the geometric size of the crystal. For highly reflective, conservative boundaries, it may be many times larger than the crystal's size. Using the crystal size should therefore be a lower bound on thermal conductivity for a sufficiently pure sample.}, $\omega$ is the angular phonon frequency, $A$ describes the strength of isotopic scattering, and $B_1$ the strength of the three-phonon scattering interactions. All of these interactions represent an inelastic scattering event which can contribute to phonon thermalization and therefore signal loss for a phonon calorimeter.

We have defined a benchmark time constant for boundary scattering ($\tau_{b}=\frac{L}{c_s}$) and the critical frequency for phonons
\begin{align}
    \omega_p = \left(\frac{c_s}{AL}\right)^{1/4} = \left(\frac{1}{A\tau_b}\right)^{1/4}
\end{align}
below which boundary scattering dominates phonon relaxation time, and above which isotopic scattering is the more important process. We also defined the three-phonon scattering frequency
\begin{equation}
\omega_B = \sqrt{\frac{1}{\tau_bB_1T^3}} = \sqrt{\frac{c_s}{LB_1T^3}}.
\end{equation}
which is explicitly temperature dependent and is assumed to be larger than $\omega_p$ here.

In a perfect crystal of finite size at low temperature, $A\rightarrow 0$ and thermal conductivity is determined entirely by crystal geometry and temperature. In particular, thermal conductivity is determined by losses at the surface (this model implicitly assumes reflections are diffusive and non-conservative, not specular). In most real crystals $A$ will be non-zero, and is thus included in the calculation as a perturbation term. In this limit, the thermal conductivity of a sample obeys the equation~\cite{Callaway}
\begin{align}
    \kappa &= \frac{2k_B\pi^2}{15}\frac{\tau_b}{c_s} \omega_T^3\left[1-\left(\frac{2\pi\omega_T}{\omega_p}\right)^4 -\frac{5}{7}\left(\frac{2\pi\omega_T}{\omega_B}\right)^2 \right] \label{eq:kappaA}
\end{align}
where $\omega_T$ is the mean phonon frequency at temperature $T$, defined as
\begin{equation}
    \omega_T = \frac{k_B T}{\hbar}.
\end{equation}
This is the mean phonon frequency at the given temperature, but to good approximation, we can use this frequency to bound the phonon energies that are expected to be limited by boundary scattering or bulk processes based on deviations from the leading order conductivity in the above equation.

Thermal conductivity measurements at low temperature  allow us to determine the purity and phonon interaction length scales for high-quality crystals, given that the volumetric specific heat is an intrinsic quantity, while the thermal conductivity depends on the extrinsic length scale of the crystal. Fitting the thermal conductivity to a model with three free parameters, $L$, $\omega_p$, and $\omega_B$, we can infer the parameters $L$, $A$, and $B_1$, assuming the sound speed and volumetric heat capacity are known for a given material.

High purity samples of Ge, Si, SiC, and diamond all demonstrate $T^3$ dependence and scaling with crystal size for temperatures below 10~K~\cite{SLACK1973321} and mm-scale crystals.  This allows us to assert that phonons with energy below 10~K have lifetime limited by boundary scattering in 1~mm size crystals for all four substrates. The scaling law here also implies that high-purity SiC crystals should have similar $\kappa_0$ to diamond, which is indeed shown to be the case in Ref.~\cite{SLACK1973321}.

To understand this scaling, we can re-write Eq.~\eqref{eq:kappaA} in terms of the specific heat of the crystal. We use the volumetric Debye specific heat,
\begin{equation}
    C_V = \frac{12\pi^4 n k_B}{5}\left(\frac{T}{T_D}\right)^3,
\end{equation}
where $k_B T_D \equiv \pi \hbar c_s ( 6 \rho/\pi)^{1/3}$ is the Debye temperature (values for $\hbar \omega_{\rm Debye} = k_B T_D$ are given in Table.~\ref{tab:properties}). This gives us the modified equation
\begin{align}
    \kappa &= \frac{1}{3}C_Vc_sL\left[1-\left(\frac{2\pi\omega_T}{\omega_p}\right)^4 -\frac{5}{7}\left(\frac{2\pi\omega_T}{\omega_B}\right)^2 \right] \\
    &= \kappa_0\left[1-\left(\frac{2\pi\omega_T}{\omega_p}\right)^4 -\frac{5}{7}\left(\frac{2\pi\omega_T}{\omega_B}\right)^2 \right] 
\end{align}
with $\kappa_0 = C_V c_s L/3$. We thus find that the temperature dependence of the leading order term is only due to the increase in the thermal phonon population, and thus dividing a measurement of thermal conductivity by heat capacity gives a temperature dependent measure of effective mean phonon lifetime ($\frac{\kappa}{C_V} = \frac{1}{3}c_s^2\tau_{b}$). 

The ultimate limit to ballistic phonon propagation for a crystal of given purity can be taken in the limit $T\rightarrow 0$, in which phonons below $\omega_p$ can be considered ballistic, and those above $\omega_p$ are unlikely to propagate to sensors at the crystal surface. Because $\omega_p$ depends on crystal purity and crystal size, we can use this property to inform our design rules. Suppose we have a superconducting phonon absorber with minimum gap frequency
\begin{equation}
    \omega_g\sim \frac{7k_bT_c}{2\hbar}.
\end{equation}
In order to achieve a high collection efficiency, we need a large crystal sufficiently pure enough to ensure that
\begin{align}
    \omega_g &\ll \omega_p \\
    \frac{7k_bT_c}{2\hbar} &\ll \left(\frac{c_s}{AL}\right)^{1/4} \\
    T_c &\ll \frac{2\hbar}{7k_b}\left(\frac{c_s}{AL}\right)^{1/4}.
\end{align}
This matches the general intuition that higher sound speed implies that the mean phonons have a higher energy, and higher-gap absorbers are acceptable. This condition allows for a quantitative crystal size optimization given known crystal purity, and allows us to compare crystals using low-temperature thermal conductance data (which can be used to extract $A$).

\section{Phonon Transmission Probabilities}\label{app:ptrans}

The phonon transmission probabilities across material interfaces are estimated using the acoustic mismatch model in \cite{Kaplan:1979}. The calculation is completely analogous to that of electromagnetic wave propagation across a boundary, except for phonons we also have an additional longitudinal mode. 

\begin{table*}[!t]
    \centering
    \begin{tabular}{|c||c|c|c||c|c|c||c|c|c||c|c|c|}
        \hline
        \bf Substrate & \multicolumn{3}{c||}{\bf Si} & \multicolumn{3}{c||}{\bf Diamond} & \multicolumn{3}{c||}{\bf 3C-SiC} & \multicolumn{3}{c||}{\bf 4H/6H-SiC}\\
        \hline
        Material  & $n_{l}$ & $n_{t}$ & $\bar{n}$ & $n_{l}$ & $n_{t}$ & $\bar{n}$ & $n_{l}$ & $n_{t}$ & $\bar{n}$ & $n_{l}$ & $n_{t}$ & $\bar{n}$ \\
        \hline 
        Al  & 0.98 & 0.89 & 0.91 & 0.72 & 0.63 & 0.65 & 0.85 & 0.95 & 0.94 & 0.86 & 0.82 & 0.83\\
        Al2O3  & 0.76 & 0.61 & 0.64 & 0.96 & 0.88 & 0.90 & 0.90 & 0.32 & 0.35 & 0.95 & 0.97 & 0.96\\
        Diamond  & 0.30 & 0.11 & 0.15 & 1.00 & 1.00 & 1.00 & 0.42 & 0.06 & 0.08 & 0.61 & 0.29 & 0.34\\
        Ga  & 0.97 & 0.89 & 0.90 & 0.67 & 0.58 & 0.61 & 0.84 & 0.90 & 0.89 & 0.84 & 0.79 & 0.80\\
        Ge  & 0.90 & 0.83 & 0.84 & 0.90 & 0.82 & 0.84 & 0.87 & 0.90 & 0.89 & 0.94 & 0.91 & 0.91\\
        In  & 0.97 & 0.89 & 0.90 & 0.68 & 0.59 & 0.61 & 0.88 & 0.90 & 0.90 & 0.86 & 0.79 & 0.80\\
        Ir  & 0.47 & 0.36 & 0.38 & 0.84 & 0.76 & 0.78 & 0.63 & 0.37 & 0.38 & 0.70 & 0.55 & 0.58\\
        Nb  & 0.83 & 0.74 & 0.75 & 0.93 & 0.85 & 0.87 & 0.92 & 0.86 & 0.86 & 0.97 & 0.88 & 0.89\\
        Si  & 1.00 & 1.00 & 1.00 & 0.79 & 0.71 & 0.73 & 0.86 & 0.48 & 0.51 & 0.89 & 0.85 & 0.85\\
        3C-SiC & 0.77 & 0.96 & 0.93 & 0.84 & 0.75 & 0.78 & 1.00 & 1.00 & 1.00 & 0.97 & 0.80 & 0.83\\
        4H/6H-SiC  & 0.58 & 0.48 & 0.50 & 0.95 & 0.86 & 0.89 & 0.71 & 0.24 & 0.27 & 1.00 & 1.00 & 1.00\\
        Sn  & 0.94 & 0.85 & 0.86 & 0.84 & 0.76 & 0.78 & 0.90 & 0.87 & 0.87 & 0.94 & 0.89 & 0.89\\
        Ta  & 0.65 & 0.54 & 0.56 & 0.95 & 0.87 & 0.89 & 0.80 & 0.58 & 0.59 & 0.88 & 0.76 & 0.78\\
        Ti  & 0.95 & 0.90 & 0.91 & 0.86 & 0.77 & 0.79 & 0.91 & 0.95 & 0.95 & 0.95 & 0.89 & 0.90\\
        W  & 0.51 & 0.41 & 0.43 & 0.88 & 0.80 & 0.82 & 0.67 & 0.42 & 0.44 & 0.75 & 0.61 & 0.64\\
        Zn  & 0.91 & 0.81 & 0.82 & 0.90 & 0.81 & 0.83 & 0.89 & 0.87 & 0.87 & 0.95 & 0.90 & 0.90\\
        \hline

    \end{tabular}
    \caption{Phonon transmission probabilities for materials relevant to the detector designs discussed in this paper. The probability $n_l$ ($n_t$) is the probability for a longitudinal (transverse) phonon incident on the interface from the substrate to get into the film. $\bar{n}$ is the probability averaged by the density of states of the two modes.}
    \label{tab:trans}
\end{table*}

\begin{figure}[t!]
\begin{center}
\includegraphics[width=0.4\textwidth]{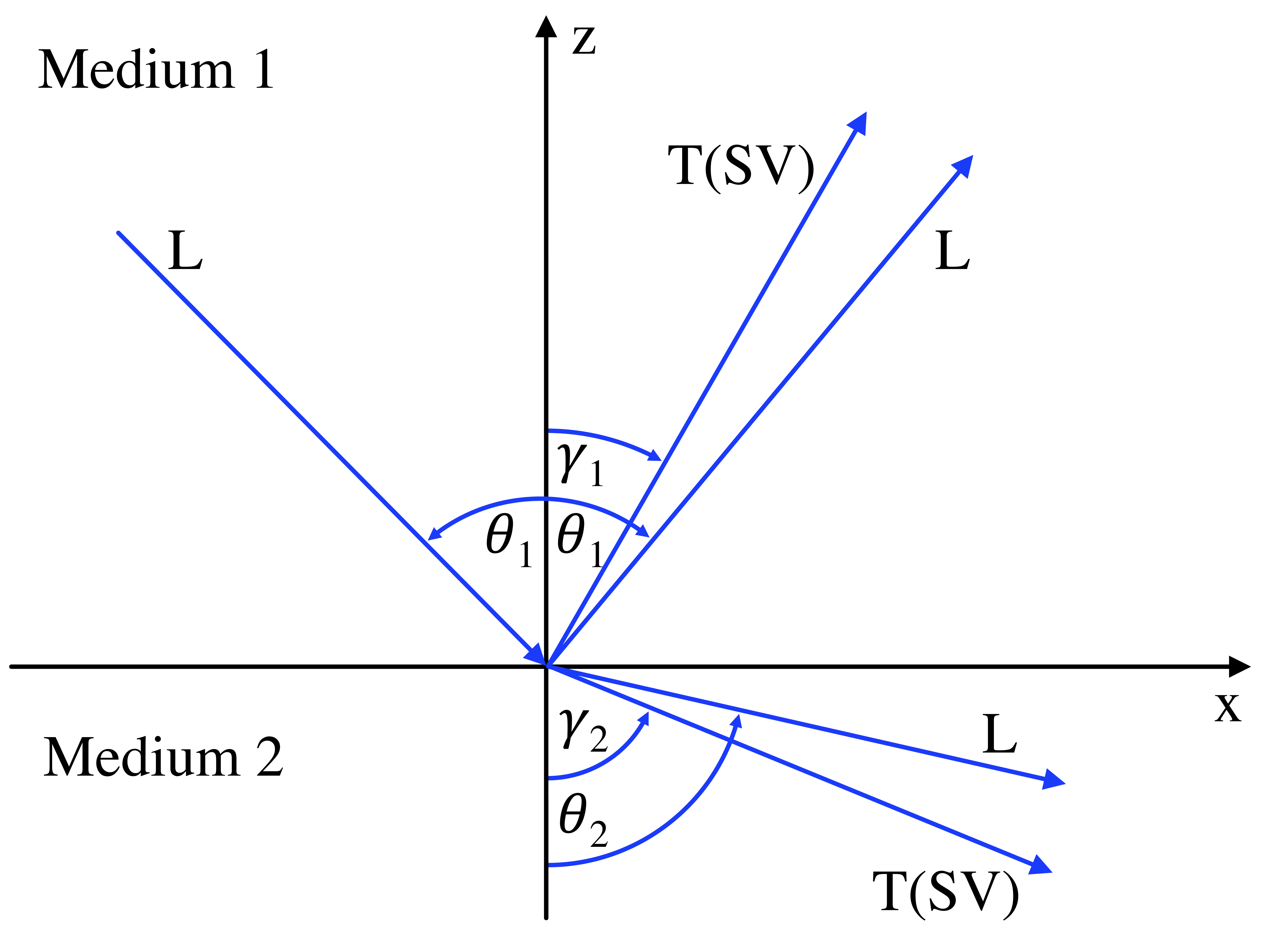}
\caption{ An incident longitudinal (L) acoustic wave is both reflected and refracted into longitudinal and transverse (SV) modes in the two media. The various angles here satisfy geometric-optical relations such as law of reflection and Snell's law. Reproduced from \cite{Kaplan:1979}. }
\label{fig:acoustic_mismatch}
\end{center}
\end{figure}

An exemplary situation is illustrated by Fig.~\ref{fig:acoustic_mismatch}. An longitudinal wave is propagating in the $x-z$ plane with the interface between medium 1 and medium 2 situated along the $x$-axis. The incoming wave can be reflected and refracted into both longitudinal and co-planar transverse mode (but not the transverse mode parallel to $y$-axis). The various angles are related via laws of geometric optics. For example, we have 
\begin{align}
    \frac{\sin \theta_1}{c_{l1}} = \frac{\sin \theta_2}{c_{l2}} = \frac{\sin \gamma_1}{c_{t1}} = \frac{\sin \gamma_2}{c_{t2}} \label{eq:snell}
\end{align}
where $c$ denotes the speed of sound with subscripts denoting the polarization and the medium. We assume isotropy and do not include any angular dependence in the speed of sounds.

In order to calculate the transmission coefficient we assume all our waves are plane waves with various amplitudes. For example the incident wave can be written as 
\begin{align}
    v_{inc} = B \exp(i {\bf k}_{inc}\cdot{\bf r} )  = B\exp( -i\beta z + i\sigma x   ) \label{eq:plane_wave}
\end{align}
where $v_{inc}$ denotes the particle velocity due to the incident acoustic wave, $B$ is the amplitude of the incident wave, $\beta =\omega/c_{l1} \cos \theta_1$ and $\sigma =\omega/c_{l1} \sin \theta_1$.

We can relate the various amplitudes using 4 boundary conditions:
\begin{enumerate}
    \item The sum of normal (tangential) components of the particle velocity at the boundary should be continuous:
    \begin{align}
        v_{1\perp} = v_{2\perp} , v_{1\parallel}=v_{2\parallel}
    \end{align}
    \item The sum of normal (tangential) components of the mechanical stress at the boundary should be continuous:
    \begin{align}
        \rho_1 c_1^2 \frac{\partial v_1}{\partial z} = \rho_2 c_2^2 \frac{\partial v_2}{\partial z}~ {\rm ,etc.}
    \end{align}
\end{enumerate}
Writing these boundary conditions and combining with Eqs.~\ref{eq:snell} and \ref{eq:plane_wave} would produce a system of linear equations for the various amplitudes which can be solved to obtain the transmission coefficient $\eta$ as a function of incident angle. The phonon transmission probability $n$ across the boundary is then defined as the angular average of the transmission coefficient:
\begin{align*}
    n &= \int_0^{\pi/2} \eta(\theta_1) \sin(2\theta) d\theta \\
    &= \int_0^{\theta_c} \eta(\theta_1) \sin(2\theta) d\theta
\end{align*}
where $\theta_c$ is the critical angle. For detailed derivation we refer reader to the appendix of Ref.~\cite{Kaplan:1979}.

Table \ref{tab:trans} contains a collection of transmission probabilities calculated in this manner.

\section{Brillouin Zones} \label{app:structurefigures}

\begin{figure}[!ht]
\centering
    \subfloat[][]{
    \includegraphics[width=0.48\linewidth]{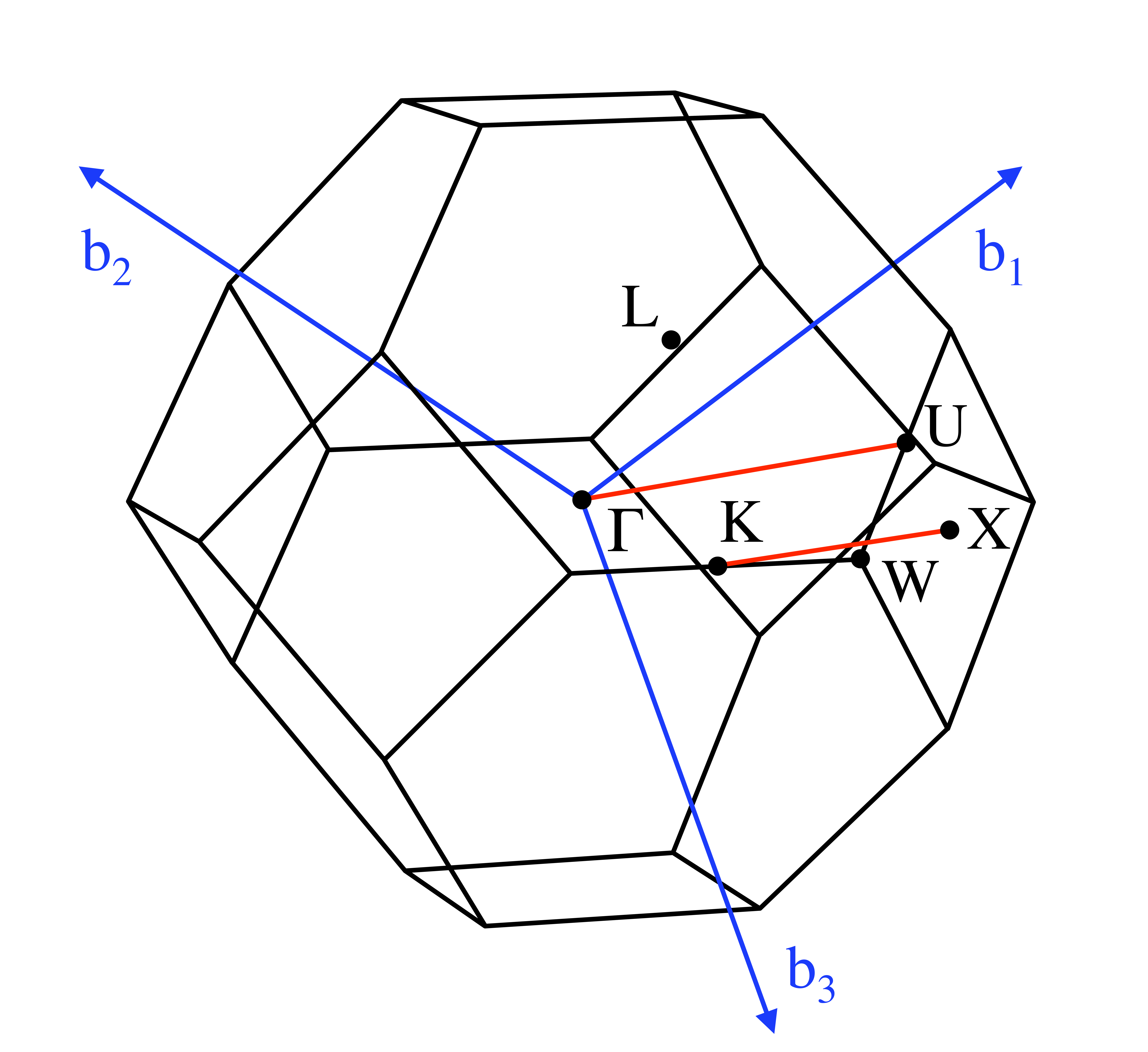}
    }
    \subfloat[][]{
    \includegraphics[width=0.48\linewidth]{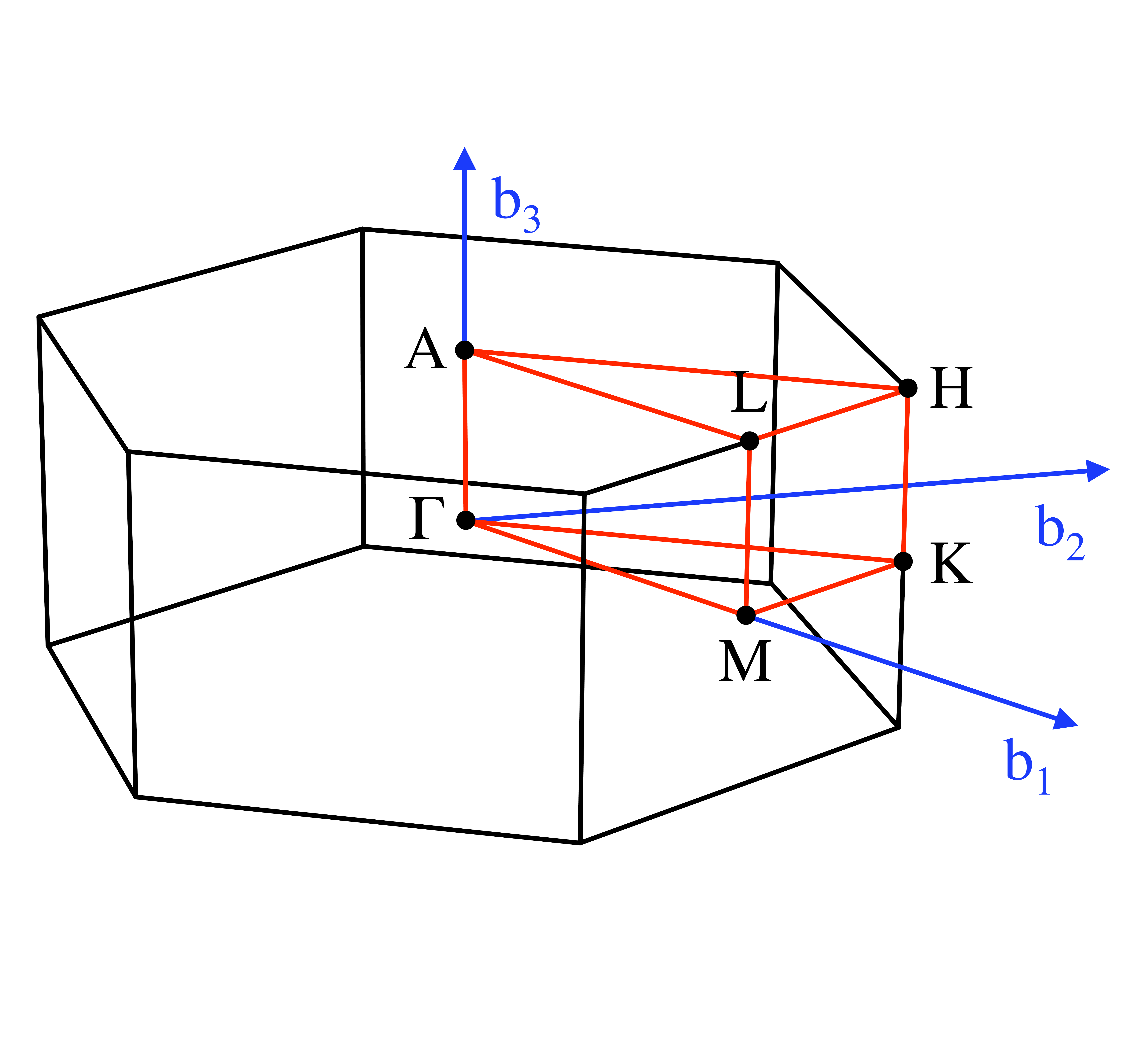}
    }
    \\[-0.5ex]
    \subfloat[][]{
    \includegraphics[width=0.5\linewidth]{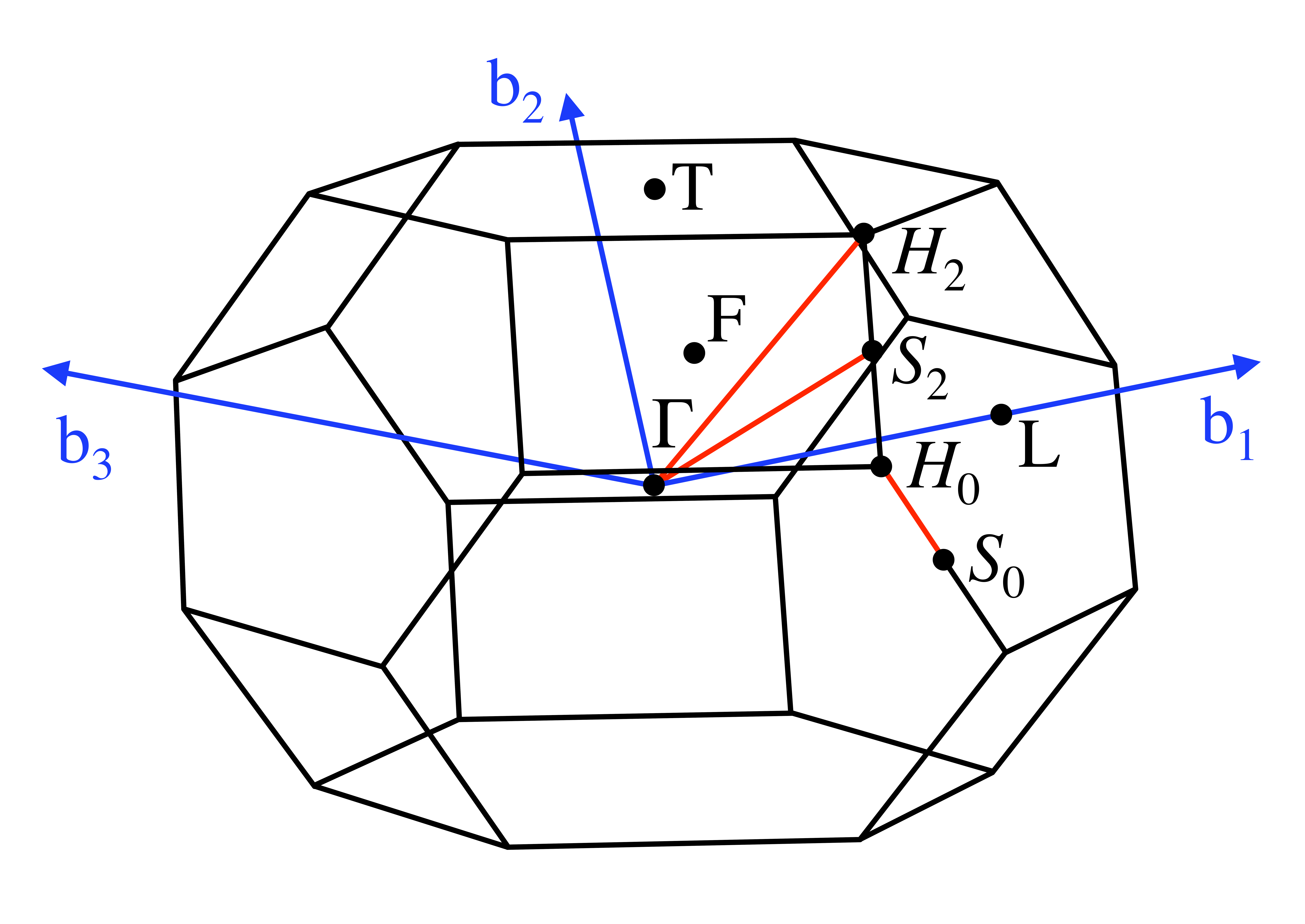}
    }
\caption{Brillouin zones and high-symmetry points for the polytypes of SiC. (a) Face-centred cubic type for the 3C polytype. (b) Primitive hexagonal type for the 2H, 4H, 6H and 8H polytypes. (c) Rhombohedral hexagonal type for the 15R polytype.}
    \label{fig:BZ}
    \end{figure}

Fig.~\ref{fig:BZ} shows the Brillouin zones for each of the lattice symmetries considered in this paper. Of particular importance are the X-valleys in the face-centered cubic type, and the L-M symmetry line in the hexagonal type, as shown in Fig.~\ref{fig:bandStructure}. These points have a basic rotational symmetry about one of the cardinal axes. The rhombohedral type is more complex, integrating a twisted crystal structure which results in asymmetric valleys. The F symmetry point is the location of the indirect gap.

\bibliographystyle{revtex-4-1.bst}
\bibliography{main}

\end{document}